# RESOURCE USAGE ANALYSIS FOR THE $\pi$-CALCULUS *

NAOKI KOBAYASHI [a], KOHEI SUENAGA [b], AND LUCIAN WISCHIK [c]

[a] 6-3-09, Aoba, Aramaki-aza, Aoba-ku, Sendai, Miyagi, Japan
*e-mail address*: koba@ecei.tohoku.ac.jp

[b] 6-3-09, Aoba, Aramaki-aza, Aoba-ku, Sendai, Miyagi, Japan
*e-mail address*: kohei@yl.is.s.u-tokyo.ac.jp

[c] 1 Microsoft Way, Redmond, WA 98052, USA
*e-mail address*: lwischik@microsoft.com

Abstract. We propose a type-based resource usage analysis for the $\pi$-calculus extended with resource creation/access primitives. The goal of the resource usage analysis is to statically check that a program accesses resources such as files and memory in a valid manner. Our type system is an extension of previous behavioral type systems for the $\pi$-calculus. It can guarantee the safety property that no invalid access is performed, as well as the property that necessary accesses (such as the close operation for a file) are eventually performed unless the program diverges. A sound type inference algorithm for the type system is also developed to free the programmer from the burden of writing complex type annotations. Based on our algorithm, we have implemented a prototype resource usage analyzer for the $\pi$-calculus. To the authors' knowledge, this is the first type-based resource usage analysis that deals with an expressive concurrent language like the $\pi$-calculus.

## 1. Introduction

Computer programs access many external resources, such as files, library functions, device drivers, etc. Such resources are often associated with certain access protocols; for example, an opened file should be eventually closed and after the file has been closed, no read/write access is allowed. The aim of resource usage analysis [11] is to statically check that programs conform to such access protocols. Although a number of approaches, including type systems and model checking, have been proposed so far for the resource usage analysis or similar analyses [5, 6, 7, 11, 1], most of them focused on analysis of sequential programs, and did not treat concurrent programs, especially those involving dynamic creation/passing of channels and resources.

In the present paper, we propose a type-based method of resource usage analysis for *concurrent languages*. Dealing with concurrency is especially important because concurrent

* A preliminary version appeared in 7th International Conference on Verification, Model Checking, and Abstract Interpretation (VMCAI 2006), Charleston, SC, USA, January 8–10, 2006.

programs are hard to debug, and also because actual programs accessing resources are often concurrent. We use the $\pi$-calculus (extended with resource primitives) as a target language so that our analysis can be applied to a wide range of concurrency primitives (including those for dynamically creating and passing channels) in a uniform manner.

A main new difficulty in dealing with concurrent programs is that control structures are more complex in concurrent programs than in sequential programs. For example, consider the following process $P_1$:

$$(\nu c)\,(\mathbf{read}(x).\overline{c}\langle\,\rangle \mid c(\,).\,\mathbf{close}(x))$$

Here, $\mathbf{read}(x).\overline{c}\langle\,\rangle$ reads $x$ and then sends a signal on channel $c$, and in parallel to that, $c(\,).\,\mathbf{close}(x)$ waits for a signal on channel $c$ and then closes $x$. Because of the synchronization through channel $c$, $x$ is closed only after being read. To capture this kind of causal dependency between communications and resource access, we use CCS processes as extra type information (which are called behavioral types). For example, the above process is given the behavioral type $(\nu c)\,(x^R.\overline{c} \mid c.\,x^C)$.

Using the behavioral types introduced above, we can construct a type system for resource usage analysis in a manner similar to previous behavioral type systems for the $\pi$-calculus [10, 3]. A type judgment is of the form $\Gamma \triangleright P : A$, where $\Gamma$ is the usual type environment and $A$ is a behavioral type approximating the behavior of $P$ on the free channels and resources. For example, the above process $P_1$ is typed $x : \mathbf{res} \triangleright P_1 : (\nu c)\,(x^R.\overline{c} \mid c.\,x^C)$. Behavioral types are also used to augment channel types. The judgment for $s(x).\,P_1$ is given by:

$$\Gamma \quad \triangleright \quad s(x).\,P_1 \quad : \quad s$$

where $\Gamma = s{:}\mathbf{chan}\langle(x{:}\mathbf{res})(\nu c)\,(x^R.\overline{c} \mid c.\,x^C)\rangle$. Here, the behavioral type of $s(x).\,P_1$ is simply a single input command $s$: the characteristic feature of this kind of type system is that the behavior of the input continuation is accounted for at output, not at input. The channel $s$ has *argument type* $(x{:}\mathbf{res})(\nu c)\,(x^R.\overline{c} \mid c.\,x^C)$, which specifies that the resource sent along channel $s$ will be read first and then closed. Using the same type environment, the output process $\overline{s}\langle r\rangle$ is typed as:

$$\Gamma, r{:}\mathbf{res} \quad \triangleright \quad \overline{s}\langle r\rangle \quad : \quad \overline{s}.\,(\nu c)\,(r^R.\overline{c} \mid c.\,r^C)$$

Here the behavioral type is an output followed by a continuation. The continuation $(\nu c)\,(r^R.\overline{c} \mid c.\,r^C)$ has been obtained by substituting $r$ for $x$ in the argument type of $s$. In this way, the types propagate information about how resources and channels passed thorough channels are accessed.

An important property of our type system is that types express abstract behavior of processes, so that certain properties of processes can be verified by verifying the corresponding properties of their types, using, for example, model checking techniques. The latter properties (of behavioral types) are more amenable to automatic verification techniques like model checking than the former ones, because the types do not have channel mobility and also because the types typically represent only the behavior of a part of the entire process.

The technical contributions of the present work are summarized as follows.

- Formalization of type systems for resource usage analysis for the $\pi$-calculus, and proof of their soundness. We have augmented previous behavioral types for the $\pi$-calculus with hiding and renaming constructors, and adapted them to the problem of resource usage analysis. CCS-like processes have been used as types also in previous work on type systems for the $\pi$-calculus [10, 3]. Igarashi and Kobayashi [10], however, used a fragment without hiding and renaming, and Chaki et al. [3] used a fragment without renaming, while the present paper uses both hiding and renaming. The inclusion of hiding and renaming is important both for accuracy and for automatic inference (see Remark 3.12).

- Realization of fully automatic verification (while making the analysis more precise than [10]). Igarashi and Kobayashi [10] gave only an abstract type system, without giving a concrete type inference algorithm. Chaki et al. [3] requires type annotations. The full automation was enabled by a combination of a number of small ideas, like inclusion of hiding and renaming as type constructors, and approximation of a CCS-like type by a Petri net (to reduce the problem of checking conformance of inferred types to resource usage specification).
- Verification of not only the usual safety property that an invalid resource access does not occur, but also an extended safety (which we call *partial liveness*) that necessary resource accesses (e.g. closing of a file) are eventually performed unless the whole process diverges. The partial liveness is not guaranteed by Chaki et al.'s type system [3]. A noteworthy point about our type system for guaranteeing the partial liveness is that it is parameterized by a mechanism that guarantees deadlock-freedom (in the sense of Kobayashi's definition [14]). So, our type system can be combined with *any* mechanism (model checking, abstract interpretation, another type system, or whatever) to verify deadlock-freedom for deadlock- or lock-freedom (e.g., Yoshida's graph type system [25]).
- Implementation of a prototype resource usage analyzer based on the proposed method. The implementation can be tested at `http://www.yl.is.s.u-tokyo.ac.jp/~kohei/usage-p`

The rest of this paper is structured as follows. Section 2 introduces an extension of the $\pi$-calculus with primitives for creating and accessing resources. Section 3 introduces a type system for resource usage analysis, which guarantees that well-typed processes never perform an invalid resource access. Section 4 gives a type inference algorithm for the type system. Section 5 extends the type system to guarantee that necessary resource accesses (such as closing of opened files) are eventually performed (unless the program diverges). Section 6 describes a prototype resource usage analyzer we have implemented based on the present work. Section 7 discusses related work. Section 8 concludes.

## 2. Processes

This section introduces the syntax and the operational semantics of our target language.

### 2.1. Syntax.

**Definition 2.1** (processes)**.** The set of processes is defined by the following syntax.

$$\begin{array}{lcl} P \text{ (processes)} & ::= & \mathbf{0} \mid \overline{x}\langle v_1,\ldots,v_n\rangle.\,P \mid x(y_1,\ldots,y_n).\,P \\ & & \mid (P \,|\, Q) \mid \textbf{if } v \textbf{ then } P \textbf{ else } Q \\ & & \mid (\nu x)\,P \mid *P \mid \mathbf{acc}_\xi(x).P \mid (\mathfrak{N}^\Phi x)P \\ v \text{ (values)} & ::= & x \mid \textbf{true} \mid \textbf{false} \end{array}$$

Here, $x$, $y$, and $z$ range over a countably infinite set **Var** of variables. $\xi$ ranges over a set of labels called *access labels*. $\Phi$, called a *trace set*, denotes a set of sequences of access labels that is prefix-closed. The prefixes (like $(\nu x)$ and $(\mathfrak{N}^\Phi x)$) bind tighter than the parallel composition $|$.

An access label specifies the kind of an access operation. Typical access labels that we are going to use in this paper are: $I$ for initialization, $R$ for read, $W$ for write, and $C$ for close.

Process $\mathbf{acc}_\xi(x).P$ accesses the resource $x$, and then behaves like $P$. We will often write $\mathbf{init}(x).P$, $\mathbf{read}(x).P$, $\mathbf{write}(x).P$, and $\mathbf{close}(x).P$ for $\mathbf{acc}_I(x).P$, $\mathbf{acc}_R(x).P$, $\mathbf{acc}_W(x).P$, $\mathbf{acc}_C(x).P$. Process $(\mathfrak{N}^\Phi x)P$ creates a new resource with the bound name $x$ that should be accessed according to $\Phi$, and then behaves like $P$. $\Phi$ specifies a set of acceptable sequences of operations that are allowed for the new resource $x$. For example, $(\mathfrak{N}^{(I(R+W)^*C)^\#} x)P$ creates a resource that should be first initialized, read or written an arbitrary number of times, and then closed. Here, $(S)^\#$ is the prefix closure of $S$, i.e., $\{s \mid ss' \in S\}$. We write $\epsilon$ for the empty sequence.

We often abbreviate a sequence $v_1, \ldots, v_n$ to $\widetilde{v}$, and write $\overline{x}\langle\widetilde{v}\rangle.\,P$ and $x(\widetilde{y}).\,P$ for $\overline{x}\langle v_1, \ldots, v_n\rangle.\,P$ and $x(y_1, \ldots, y_n).\,P$. We often omit trailing $\mathbf{0}$ and write $\overline{x}\langle\widetilde{v}\rangle$ and $\mathbf{acc}_\xi(x)$ for $\overline{x}\langle\widetilde{v}\rangle.\,\mathbf{0}$ and $\mathbf{acc}_\xi(x).\mathbf{0}$ respectively.

The bound and free variables of $P$ are defined in a customary manner; also $(\mathfrak{N}^\Phi x)P$ binds $x$. We identify processes up to $\alpha$-conversion, and assume that $\alpha$-conversion is implicity applied so that bound variables are always different from each other and from free variables.

2.2. **Operational Semantics.** We now formally define the operational semantics of our process calculus The operational semantics is almost the same as the standard reduction semantics for the $\pi$-calculus, except that trace sets $\Phi$ (which represent how resources should be accessed in future) may change during reduction.

**Definition 2.2.** The *structural preorder* $\preceq$ is the least reflexive and transitive relation closed under the rules in Figure 1 ($P \equiv Q$ stands for $(P \preceq Q) \wedge (Q \preceq P)$).

**Remark 2.3.** As in our previous behavioural type systems for the $\pi$-calculus [10, 14, 15], the structural relation is asymmetric. If the standard, symmetric structural relation were used, the type preservation property would not hold: $\Gamma \triangleright *P \,|\, P : A$ does not necessarily imply $\Gamma \triangleright *P : A$) for the type system introduced in the next section.

**Definition 2.4.** The set of *reduction labels*, ranged over by $L$, is $\{x^\xi \mid x \in \mathbf{Var}\} \cup \{\tau\}$. We define *target*$(L)$ by:

$$\mathit{target}(x^\xi) = \{x\} \qquad \mathit{target}(\tau) = \emptyset$$

**Definition 2.5.** Let $\Phi$ be a set of sequences of access labels. $\Phi^{-\xi}$ is defined by: $\Phi^{-\xi} = \{s \mid \xi s \in \Phi\}$.

**Definition 2.6.** The reduction relation $\xrightarrow{L}$ is the least relation closed under the rules in Figure 2.

We write $P \longrightarrow Q$ when $P \xrightarrow{L} Q$ for some $L$. We write $\longrightarrow^*$ for the reflexive and transitive closure of $\longrightarrow$.

Notice that when an invalid access to a resource occurs (i.e. when the program accesses $\xi$ but the specification $\Phi$ has no $\xi$-prefixes), then resource specification $\Phi$ is set to $\emptyset$ by

$$P \mid \mathbf{0} \equiv P \qquad \text{(SP-Zero)}$$

$$P \mid Q \equiv Q \mid P \qquad \text{(SP-Commut)}$$

$$P \mid (Q \mid R) \equiv (P \mid Q) \mid R \qquad \text{(SP-Assoc)}$$

$$*P \preceq *P \mid P \qquad \text{(SP-Rep)}$$

$$(\nu x)\, P \mid Q \preceq (\nu x)\, (P \mid Q) (\text{if } x \text{ not free in } Q) \qquad \text{(SP-New)}$$

$$(\mathfrak{N}^{\Phi} x)P \mid Q \preceq (\mathfrak{N}^{\Phi} x)(P \mid Q) (\text{if } x \text{ not free in } Q) \qquad \text{(SP-NewR)}$$

$$\frac{P \preceq P' \qquad Q \preceq Q'}{P \mid Q \preceq P' \mid Q'} \qquad \text{(SP-Par)}$$

$$\frac{P \preceq Q}{(\nu x)\, P \preceq (\nu x)\, Q} \qquad \text{(SP-CNew)}$$

$$\frac{P \preceq Q}{(\mathfrak{N}^{\Phi} x)P \preceq (\mathfrak{N}^{\Phi} x)Q} \qquad \text{(SP-CNewR)}$$

Figure 1: Structural Preorder

$$\overline{x}\langle \widetilde{z} \rangle.\, P \mid x(\widetilde{y}).\, Q \xrightarrow{\tau} P \mid [\widetilde{z}/\widetilde{y}]Q \qquad \text{(R-Com)}$$

$$\frac{P \xrightarrow{L} Q \qquad x \notin \mathit{target}(L)}{(\nu x)\, P \xrightarrow{L} (\nu x)\, Q} \qquad \text{(R-New)}$$

$$\mathbf{acc}_{\xi}(x).P \xrightarrow{x^{\xi}} P \qquad \text{(R-Acc)}$$

$$\frac{P \xrightarrow{x^{\xi}} Q}{(\mathfrak{N}^{\Phi} x)P \xrightarrow{\tau} (\mathfrak{N}^{\Phi^{-\xi}} x)Q} \qquad \text{(R-NewR1)}$$

$$\frac{P \xrightarrow{L} Q}{P \mid R \xrightarrow{L} Q \mid R} \qquad \text{(R-Par)}$$

$$\frac{P \xrightarrow{L} Q \qquad x \notin \mathit{target}(L)}{(\mathfrak{N}^{\Phi} x)P \xrightarrow{L} (\mathfrak{N}^{\Phi} x)Q} \qquad \text{(R-NewR2)}$$

$$\textbf{if true then } P \textbf{ else } Q \xrightarrow{\tau} P \qquad \text{(R-IfT)}$$

$$\textbf{if false then } P \textbf{ else } Q \xrightarrow{\tau} Q \qquad \text{(R-IfF)}$$

$$\frac{P \preceq P' \quad P' \xrightarrow{L} Q' \quad Q' \preceq Q}{P \xrightarrow{L} Q} \qquad \text{(R-SP)}$$

Figure 2: Reduction Relation

(R-NewR1). On the other hand $\Phi \supseteq \{\epsilon\}$ indicates a resource that has been correctly used so far, and $\Phi = \{\epsilon\}$ indicates one that has been correctly and completely used.

**Definition 2.7.** A process $P$ is *resource-safe* if it does not contain a sub-expression of the form $(\mathfrak{N}^{\emptyset} x)Q$.

We give a type system guaranteeing that any resource-safe, well-typed process cannot be reduced to a non-safe process (in other words, any resource-safe, well-typed process never performs an invalid access) in Section 3.

**Example 2.8.** The following process first creates a resource $x$ that should be first initialized, read an arbitrary number of times, and then closed. It then spawns four processes; they synchronize through channels $c_1$ and $c_2$, so that $x$ is accessed in a valid order.

$$
\begin{array}{ll}
(\mathfrak{N}^{(IR^*C)^\#} x)(\nu c_1)\,(\nu c_2)\,\big( & \\
\quad \mathbf{init}(x).(\overline{c_1}\langle\,\rangle \,|\, \overline{c_1}\langle\,\rangle) & /\text{* } \textit{initialize } x\textit{, and send signals} \text{ */} \\
\quad |\, c_1(\,).\,\mathbf{read}(x).\overline{c_2}\langle\,\rangle & /\text{* } \textit{wait for a signal on } c_1\textit{,} \\
 & \quad \textit{then read } x\textit{, and signal on } c_2 \text{ */} \\
\quad |\, c_1(\,).\,\mathbf{read}(x).\overline{c_2}\langle\,\rangle & /\text{* } \textit{wait for a signal on } c_1\textit{,} \\
 & \quad \textit{then read } x\textit{, and signal on } c_2 \text{ */} \\
\quad |\, c_2(\,).\,c_2(\,).\,\mathbf{close}(x)\,\big) & /\text{* } \textit{wait on } c_2\textit{, then close } x \text{ */}
\end{array}
$$

□

**Example 2.9.** The following program is prototypical of recursive functions. There is a replicated service which listens on channel $s$; it either terminates the recursion by sending a message back on the reply channel $r$, or it recursively invokes a sub-instance of itself which will reply on a private channel $r'$. In this example each recursive step does a $\mathbf{read}(x)$. The following program use an integer to decide whether or not to recurse. Though our language does not have integers and operations on them as primitives, it is trivial to extend our language and type system with those primitives.

$$
\begin{array}{l}
(\nu s)\,\big(\,*(s(n,x,r).\,\mathbf{if}\ n=0\ \mathbf{then}\ \overline{r}\langle\rangle \\
\qquad\qquad \mathbf{else}\ (\nu r')\,(\overline{s}\langle n-1,x,r'\rangle \,|\, r'().\,\mathbf{read}(x).\overline{r}\langle\rangle) \\
\quad |\ \ (\mathfrak{N}^{(IR^*C)^\#} x)(\nu r)\,(\mathbf{init}(x).\overline{s}\langle 100,x,r\rangle \,|\, r().\,\mathbf{close}(x))\,\big)
\end{array}
$$

□ The above program corresponds to the following higher-level program:

```
init(x); parbegin read(x); read(x) parend; close(x)
```

**Example 2.10.** Consider the following producer/consumer program:[1]

$$
\begin{array}{l}
(\nu \mathit{producer})\,(\nu \mathit{consumer}) \\
*(\mathit{producer}(b,p,c).\,p().\,\mathbf{acc_P}(b).(\overline{c}\langle\rangle \,|\, \overline{\mathit{producer}}\langle b,p,c\rangle)) \,| \\
*(\mathit{consumer}(b,p,c).\,c().\,\mathbf{acc_G}(b).(\overline{p}\langle\rangle \,|\, \overline{\mathit{producer}}\langle b,p,c\rangle)) \,| \\
(\mathfrak{N}^{((\mathbf{P}\ \mathbf{G})^*)^\#} \mathit{buf})(\nu x)\,(\nu y) \\
\quad *(\overline{\mathit{producer}}\langle \mathit{buf},x,y\rangle) \,|\, *(\overline{\mathit{consumer}}\langle \mathit{buf},x,y\rangle) \,|\, \overline{x}\langle\rangle
\end{array}
$$

The first two processes $*(\mathit{producer}(b,p,c).\ \cdots)$ and $*(\mathit{consumer}(b,p,c).\ \cdots)$ define the behavior of producers and consumers. A producer repeatedly waits to receive a signal on $p$, performs a put on the buffer $b$ (by $\mathbf{acc_P}(b)$), and then sends a signal on $c$. A consumer repeatedly waits to receive a signal on $c$, performs a get on the buffer $b$ (by $\mathbf{acc_P}(b)$), and then sends a signal on $p$. The third process creates a new buffer on which put and get should be applied only alternately, creates two channels $x$ and $y$ used for synchronization, and runs infinitely many producers and consumers.

**Remark 2.11.** We treat resources as primitives in this paper, but we could alternatively express a resource as a tuple of channels, each of which corresponds to each access operation. For example, the resource in Example 2.8 can be expressed as a tuple consisting of three channels **init**, **read**, and **close**. If we did so, we could directly reuse the previous type systems [10, 3] to infer some of the properties discussed in this paper (with different precision). Treating resources as primitives, however, simplifies the type systems introduced in later sections and clarifies the essence: if we expressed a resource as a tuple of channels, we would need primitives for simultaneous creation of multiple channels as in [10], and need to care about whether communications on the resource access channels succeed or not. On the other hand, our resource access primitives are non-blocking, which simplifies in particular the extended type system discussed in Section 5.

[1] This is an example taken from an ealier version of [20] and modified.

## 3. Type System

This section introduces a type system that prevents invalid access to resources. The type system in this section does not guarantee a liveness property that all the necessary accesses are eventually made; extensions to guarantee that property are discussed in Section 5.

3.1. **Types.** We first introduce the syntax of types. We use two categories of types: value types and behavioral types. The latter describes how a process accesses resources and communicates through channels. As mentioned in Section 1, we use CCS processes for behavioral types.

**Definition 3.1** (types)**.** The sets of *value types* $\sigma$ and *behavioral types* $A$ are defined by:

$$\begin{array}{lll} \sigma & ::= & \mathbf{bool} \mid \mathbf{res} \mid \mathbf{chan}\langle(x_1:\sigma_1,\dots,x_n:\sigma_n)A\rangle \\ A & ::= & \mathbf{0} \mid \alpha \mid a.A \mid x^{\xi}.A \mid \tau.A \mid (A_1 \mid A_2) \mid A_1 \oplus A_2 \mid *A \\ & & \mid \langle y_1/x_1,\dots,y_n/x_n\rangle A \mid (\nu x)\, A \mid \mu\alpha.A \mid A{\uparrow_S} \mid A{\downarrow_S} \\ \multicolumn{3}{l}{a \text{ (communication labels) } ::= x \mid \overline{x}} \end{array}$$

A behavioral type $A$, which is a CCS process, describes what kind of communication and resource access a process may perform. $\mathbf{0}$ describes a process that performs no communication or resource access. The types $x.\,A$, $\overline{x}.\,A$, $x^{\xi}.A$ and $\tau.A$ describe processes that first perform an action and then behave according to $A$; the actions are, respectively, an input on $x$, an output on $x$, an access operation $\xi$ on $x$, and the invisible action. $A_1 \mid A_2$ describes a process that performs communications and resource access according to $A_1$ and $A_2$, possibly in parallel. $A_1 \oplus A_2$ describes a process that behaves according to either $A_1$ or $A_2$. $*A$ describes a process that behaves like $A$ an arbitrary number of times, possibly in parallel. $\langle y_1/x_1,\dots,y_n/x_n\rangle A$, abbreviated to $\langle\widetilde{y}/\widetilde{x}\rangle A$, denotes simultaneous renaming of $\widetilde{x}$ with $\widetilde{y}$ in $A$. $(\nu x)\, A$ describes a process that behaves like $A$ for some hidden channel $x$. For example, $(\nu x)\,(x.\,\overline{y} \mid \overline{x})$ describes a process that performs an output on $y$ after the invisible action on $x$. The type $\mu\alpha.A$ describes a process that behaves like a recursive process defined by $\alpha \stackrel{\triangle}{=} A$.[2] The type $A{\uparrow_S}$ describes a process that behaves like $A$, except that actions whose targets are in $S$ are replaced by the invisible action $\tau$, while $A{\downarrow_S}$ describes a process that behaves like $A$, except that actions whose targets are not in $S$ are replaced by $\tau$. The formal semantics of behavioral types is defined later using labeled transition semantics.

As for value types, **bool** is the type of booleans. **res** is the type of resources. The type $\mathbf{chan}\langle(x_1:\sigma_1,\dots,x_n:\sigma_n)A\rangle$, abbreviated to $\mathbf{chan}\langle(\widetilde{x}:\widetilde{\sigma})A\rangle$, describes channels carrying tuples consisting of values of types $\sigma_1,\dots,\sigma_n$. Here the type $A$ approximates how a receiver on the channel may use the elements $x_1, \dots, x_n$ of each tuple for communications and resource access. For example, $\mathbf{chan}\langle(x:\mathbf{res}, y:\mathbf{res})x^R.y^C\rangle$ describes channels carrying a pair of resources, where a party who receives the actual pair $(x', y')$ will first read $x'$ and then close $y'$. We sometimes omit $\widetilde{\sigma}$ and write $\mathbf{chan}\langle(\widetilde{x})A\rangle$ for $\mathbf{chan}\langle(\widetilde{x}:\widetilde{\sigma})A\rangle$. When $\widetilde{x}$ is empty, we also write $\mathbf{chan}\langle\rangle$.

[2]The replication $*A$ and $\mu\alpha.(A \mid \alpha)$ have the same semantics in this section, but they are differentiated in Section 5 by the predicate *disabled*.

Note that $\langle\widetilde{y}/\widetilde{x}\rangle$ is treated as a constructor rather than an operator for performing the actual substitution. We write $[\widetilde{y}/\widetilde{x}]$ for the latter throughout this paper. $\langle\widetilde{y}/\widetilde{x}\rangle A$ is slightly different from the *relabeling* of the standard CCS [19]: $\langle y/x\rangle(x \mid \overline{y})$ allows the communication on $y$, but the relabeling of CCS does not. This difference calls for the introduction of a special transition label $\{x, \overline{y}\}$ in Section 3.2.

**Definition 3.2.** The set of *free variables* of $A$, written $\mathbf{FV}(A)$, is defined by:

$$\begin{array}{rcl}
\mathbf{FV}(\mathbf{0}) & = & \emptyset \\
\mathbf{FV}(\alpha) & = & \emptyset \\
\mathbf{FV}(x.\,A) & = & \{x\} \cup \mathbf{FV}(A) \\
\mathbf{FV}(\overline{x}.\,A) & = & \{x\} \cup \mathbf{FV}(A) \\
\mathbf{FV}(x^{\xi}.A) & = & \{x\} \cup \mathbf{FV}(A) \\
\mathbf{FV}(\tau.A) & = & \mathbf{FV}(A) \\
\mathbf{FV}(A_1 \mid A_2) & = & \mathbf{FV}(A_1) \cup \mathbf{FV}(A_2) \\
\mathbf{FV}(A_1 \oplus A_2) & = & \mathbf{FV}(A_1) \cup \mathbf{FV}(A_2) \\
\mathbf{FV}(*A) & = & \mathbf{FV}(A) \\
\mathbf{FV}(\langle\widetilde{y}/\widetilde{x}\rangle A) & = & (\mathbf{FV}(A)\backslash\{\widetilde{x}\}) \cup \{\widetilde{y}\} \\
\mathbf{FV}((\nu x)\,A) & = & \mathbf{FV}(A)\backslash\{x\} \\
\mathbf{FV}(\mu\alpha.A) & = & \mathbf{FV}(A) \\
\mathbf{FV}(A{\uparrow_S}) & = & \mathbf{FV}(A)\backslash S \\
\mathbf{FV}(A{\downarrow_S}) & = & \mathbf{FV}(A) \cap S
\end{array}$$

As defined above, $(\nu x)\,A$, $\langle\widetilde{y}/\widetilde{x}\rangle A$, and $A{\uparrow_S}$ bind $x$, $\widetilde{x}$, and the variables in $S$ respectively. We identify behavioral types up to renaming of bound variables. In the rest of this paper, we require that every channel type $\mathbf{chan}\langle(x_1:\sigma_1, \dots, x_n:\sigma_n)A\rangle$ must satisfy $\mathbf{FV}(A) \subseteq \{x_1, \dots, x_n\}$. For example, $\mathbf{chan}\langle(x{:}\mathbf{res})x^R\rangle$ is a valid type but $\mathbf{chan}\langle(x{:}\mathbf{res})y^R\rangle$ is not.[3]

3.2. **Semantics of behavioral types.** We give a labeled transition relation $\xrightarrow{l}$ for behavioral types. The transition labels $l$ (distinct from the reduction labels $L$ of Definition 2.4) are

$$l \ ::= \ x \mid \overline{x} \mid x^{\xi} \mid \tau \mid \{x, \overline{y}\}$$

The label $\{x, \overline{y}\}$ indicates the potential to react in the presence of a substitution that identifies $x$ and $y$. We also extend *target* to the function on transition labels by:

$$\mathit{target}(x) = \mathit{target}(\overline{x}) = \{x\} \qquad \mathit{target}(\{x, \overline{y}\}) = \{x, y\}$$

The transition relation $\xrightarrow{l}$ on behavioral types is the least relation closed under the rules in Figure 3. We write $\Longrightarrow$ for the reflexive and transitive closure of $\xrightarrow{\tau}$. We also write $\stackrel{l}{\Longrightarrow}$ for $\Longrightarrow\xrightarrow{l}\Longrightarrow$.

[3] This constraint can be removed if we assume that the free variables in *codom*$(\Gamma)$ never clash with the bound variables of $P$ in the judgment form $\Gamma \triangleright P : A$ given later. In particular, we need an implicit assumption $\{\widetilde{y}\} \cap \mathbf{FV}(\Gamma) = \emptyset$ in Figure 4, (T-In).

$$a.A \xrightarrow{a} A \qquad x^{\xi}.A \xrightarrow{x^{\xi}} A \qquad \tau.A \xrightarrow{\tau} A \qquad \text{(TR-ACT)}$$

$$\frac{A_1 \xrightarrow{l} A_1'}{A_1|A_2 \xrightarrow{l} A_1'|A_2} \qquad \frac{A_2 \xrightarrow{l} A_2'}{A_1|A_2 \xrightarrow{l} A_1|A_2'} \qquad \text{(TR-PAR1)}$$

$$\frac{A_1 \xrightarrow{x} A_1' \quad A_2 \xrightarrow{\overline{y}} A_2'}{A_1|A_2 \xrightarrow{\{x,\overline{y}\}} A_1'|A_2'} \qquad \frac{A_1 \xrightarrow{\overline{y}} A_1' \quad A_2 \xrightarrow{x} A_2'}{A_1|A_2 \xrightarrow{\{x,\overline{y}\}} A_1'|A_2'} \qquad \text{(TR-PAR2)}$$

$$\frac{A \xrightarrow{\{x,\overline{x}\}} A'}{A \xrightarrow{\tau} A'} \qquad \text{(TR-COM)}$$

$$\frac{A_1 \xrightarrow{l} A_1'}{A_1 \oplus A_2 \xrightarrow{l} A_1'} \qquad \frac{A_2 \xrightarrow{l} A_2'}{A_1 \oplus A_2 \xrightarrow{l} A_2'} \qquad \text{(TR-OR)}$$

$$\frac{A \,|\, {*A} \xrightarrow{l} A'}{{*A} \xrightarrow{l} A'} \qquad \text{(TR-REP)}$$

$$\frac{[\mu\alpha.A/\alpha]A \xrightarrow{l} A'}{\mu\alpha.A \xrightarrow{l} A'} \qquad \text{(TR-REC)}$$

$$\frac{A \xrightarrow{l} A'}{\langle\widetilde{y}/\widetilde{x}\rangle A \xrightarrow{[\widetilde{y}/\widetilde{x}]l} \langle\widetilde{y}/\widetilde{x}\rangle A'} \qquad \text{(TR-RENAME)}$$

$$\frac{A \xrightarrow{l} A' \quad \mathit{target}(l) \cap \{x\} = \emptyset}{(\nu x)\, A \xrightarrow{l} (\nu x)\, A'} \qquad \text{(TR-HIDING)}$$

$$\frac{A \xrightarrow{l} A' \quad \mathit{target}(l) \subseteq S}{A{\uparrow_S} \xrightarrow{\tau} A'{\uparrow_S}} \qquad \frac{A \xrightarrow{l} A' \quad \mathit{target}(l) \cap S = \emptyset}{A{\uparrow_S} \xrightarrow{l} A'{\uparrow_S}} \qquad \text{(TR-EXCLUDE)}$$

$$\frac{A \xrightarrow{l} A' \quad \mathit{target}(l) \subseteq S}{A{\downarrow_S} \xrightarrow{l} A'{\downarrow_S}} \qquad \frac{A \xrightarrow{l} A' \quad \mathit{target}(l) \cap S = \emptyset}{A{\downarrow_S} \xrightarrow{\tau} A'{\downarrow_S}} \qquad \text{(TR-PROJECT)}$$

Figure 3: Transition semantics of behavioral types

**Remark 3.3.** $(\nu x)\, A$ should not be confused with $A{\uparrow_{\{x\}}}$. $(\nu x)\, A$ is the hiding operator of CCS, while $A{\uparrow_{\{x\}}}$ just replaces any actions on $x$ with $\tau$ [10]. For example, $(\nu x)\,(x.\, y^{\xi})$ cannot make any transition, but $(x.\, y^{\xi}){\uparrow_{\{x\}}} \xrightarrow{\tau}\xrightarrow{y^{\xi}} \mathbf{0}{\uparrow_{\{x\}}}$.

The set $\mathbf{traces}_x(A)$ defined below is the set of possible access sequences on $x$ described by $A$.

**Definition 3.4** (traces)**.**

$$\mathbf{traces}_x(A) = \{\xi_1 \dots \xi_n \mid A{\downarrow_{\{x\}}} \stackrel{x^{\xi_1}}{\Longrightarrow} \cdots \stackrel{x^{\xi_n}}{\Longrightarrow} A'\}$$

Note that $\mathbf{traces}_x(A)$ is prefix-closed (hence a trace set) by definition.

We define the subtyping relation $A_1 \leq A_2$ below. Intuitively, $A_1 \leq A_2$ means that a process behaving according to $A_1$ can also be viewed as a process behaving according to $A_2$. To put in another way, $A_1 \leq A_2$ means that $A_2$ simulates $A_1$.We define $\leq$ for only *closed* types, i.e., those not containing free type variables.

**Definition 3.5** (subtyping)**.** The subtyping relation $\leq$ on closed behavioral types is the largest relation such that $A_1 \leq A_2$ and $A_1 \xrightarrow{l} A_1'$ implies $A_2 \stackrel{l}{\Longrightarrow} A_2'$ and $A_1' \leq A_2'$ for some $A_2'$.

We often write $A_1 \geq A_2$ for $A_2 \leq A_1$, and write $A_1 \approx A_2$ for $A_1 \leq A_2 \wedge A_2 \leq A_1$.

**Remark 3.6.** Note that the subtyping relation defined here is the converse of the one used in Igarashi and Kobayashi's generic type system [10]. This is due to two different, dual views on behavioral types. Here, we think of behavioral types as describing the behavior of processes. On the other hand, Igarashi and Kobayashi [10] think of behavioral types as describing the assumption on the environment about what kind of process is accepted by the environment. Because of this difference, they write behavioral types on the lefthand side of $\triangleright$, and write $A_1 \& A_2$ for non-deterministic choice instead of $A_1 \oplus A_2$.

**Remark 3.7.** Depending on what property the type system should guarantee, a finer subtyping relation may need to be chosen. For example, the above definition allows $(x^W.\mathbf{0}) \,|\, (x^W.\mathbf{0}) \leq x^W.x^W.\mathbf{0}$. We may want to disallow this relation if we want to infer a property like "no simultaneous writes on $x$ can occur."

The following properties are satisfied by $\leq$. For proofs, see Appendix A.

**Lemma 3.8.**

(1) *$\leq$ is a precongruence, i.e., $\leq$ is closed under any behavioral type constructor.*
(2) *If $A_1 \leq A_2$, then $\boldsymbol{traces}_x(A_1) \subseteq \boldsymbol{traces}_x(A_2)$ for any $x$.*
(3) *$B_1 \oplus B_2 \leq A$ if and only if $B_1 \leq A$ and $B_2 \leq A$ .*
(4) *If $[B/\alpha]A \leq B$, then $\mu\alpha.A \leq B$.*

3.3. **Typing.** We consider two kinds of judgments, $\Gamma \triangleright v : \sigma$ for values, and $\Gamma \triangleright P : A$ for processes. $\Gamma$ is a mapping from a finite set of variables to value types. In $\Gamma \triangleright P : A$, the type environment $\Gamma$ describes the types of the variables, and $A$ describes the possible behaviors of $P$. For example, $x : \mathbf{chan}\langle(b : \mathbf{bool})\mathbf{0}\rangle \triangleright P : \overline{x} \,|\, \overline{x}$ implies that $P$ may send booleans along the channel $x$ twice. The judgment $y : \mathbf{chan}\langle(x : \mathbf{chan}\langle(b : \mathbf{bool})\mathbf{0}\rangle)\overline{x}\rangle \triangleright Q : y$ means that $Q$ may perform an input on $y$ once, and then it may send a boolean on the received value. Note that in the judgment $\Gamma \triangleright P : A$, the type $A$ is an approximation of the behavior of $P$ on free channels. $P$ may do less than what is specified by $A$, but must not do more; for example, $x : \mathbf{chan}\langle(\,)\mathbf{0}\rangle \triangleright \overline{x}\langle\,\rangle : \overline{x} \,|\, \overline{x}$ holds but $x : \mathbf{chan}\langle(\,)\mathbf{0}\rangle \triangleright \overline{x}\langle\,\rangle.\,\overline{x}\langle\,\rangle : \overline{x}$ does not. Because of this invariant, if $A$ does not perform any invalid access, neither does $P$.

We write $dom(\Gamma)$ for the domain of $\Gamma$. We write $\emptyset$ for the empty type environment, and write $x_1 : \tau_1, \ldots, x_n : \tau_n$ (where $x_1, \ldots, x_n$ are distinct from each other) for the type environment $\Gamma$ such that $dom(\Gamma) = \{x_1, \ldots, x_n\}$ and $\Gamma(x_i) = \tau_i$ for each $i \in \{1, \ldots, n\}$. When $x \notin dom(\Gamma)$, we write $\Gamma, x : \tau$ for the type environment $\Delta$ such that $dom(\Delta) = dom(\Gamma) \cup \{x\}$, $\Delta(x) = \tau$, and $\Delta(y) = \Gamma(y)$ for $y \in dom(\Gamma)$. We define the *value judgment* relation $\Gamma \triangleright v{:}\sigma$ to be the least relation closed under

$$\Gamma,\, x{:}\sigma \triangleright x{:}\sigma \qquad \Gamma \triangleright \mathbf{true}{:}\mathbf{bool} \qquad \Gamma \triangleright \mathbf{false}{:}\mathbf{bool}.$$

We write $\Gamma \triangleright \widetilde{v}{:}\widetilde{\sigma}$ as an abbreviation for $(\Gamma \triangleright v_1{:}\sigma_1) \wedge \cdots \wedge (\Gamma \triangleright v_n{:}\sigma_n)$.

**Definition 3.9.** The type judgment relation $\Gamma \triangleright P : A$ is the least relation closed under the rules given in Figure 4.

We explain key rules below.

In rule (T-Out), the first premise $\Gamma \triangleright P : A_2$ implies that the continuation of the output process behaves like $A_2$, and the second premise $\Gamma \triangleright x : \mathbf{chan}\langle(\widetilde{y} : \widetilde{\sigma})A_1\rangle$ implies that the tuple of values $\widetilde{v}$ being sent may be used by an input process according to $\langle\widetilde{v}/\widetilde{y}\rangle A_1$. Therefore, the whole behavior of the output process is described by $\overline{x}.\,(\langle\widetilde{v}/\widetilde{y}\rangle A_1 \mid A_2)$. Here, $\langle v_1/x_1, \ldots, v_n/x_n\rangle A$ stands for $\langle v_{i_1}/x_{i_1}, \ldots, v_{i_k}/x_{i_k}\rangle A$ where $\{v_{i_1}, \ldots, v_{i_k}\} = \{v_1, \ldots, v_n\}\backslash\{\mathbf{true}, \mathbf{false}\}$. For example,$\langle \mathbf{true}/x, y/z\rangle A$ stands for $\langle y/z\rangle A$. Note that, as in previous behavioral type systems [10, 3], the resource access and communications made on $\widetilde{v}$ by the receiver of $\widetilde{v}$ are counted as the behavior of the output process (see Remark 3.13).

In rule (T-In), the first premise implies that the continuation of the input process behaves like $A_2$. Following previous behavioral type systems [10, 3], we split $A_2$ into two parts: $A_2{\downarrow}_{\{\widetilde{y}\}}$ and $A_2{\uparrow}_{\{\widetilde{y}\}}$. The first part describes the behavior on the received values $\widetilde{y}$ and is taken into account in the channel type. The second part describes the resource access and communications performed on other values, and is taken into account in the behavioral type of the input process. The condition $A_2{\downarrow}_{\{\widetilde{y}\}} \leq A_1$ requires that the access and communication behavior on $\widetilde{y}$ conforms to $A_1$, the channel arguments' behavior.

In (T-New), the premise implies that $P$ behaves like $A$, so that $(\nu x)\,P$ behaves like $(\nu x)\,A$. Here, we only require that $x$ is a channel, unlike in the previous behavioral type systems for the $\pi$-calculus [10, 15]. That is because we are only interested in the resource access behavior; the communication behavior is used only for accurately inferring the resource access behavior.

In (T-NewR), we check that the process's behavior $A$ conforms to the resource usage specification $\Phi$.

Rule (T-Sub) allows the type $A'$ of a process to be replaced by its approximation $A$.

We remark that weakening of $\Gamma$ can be derived (Appendix B, Lemma B.1) and so is not needed as a rule.

The following example shows how information about the usage of resources by an input process is propagated to an output process.

**Example 3.10.** Let us consider $(\mathfrak{N}^{\Phi}x)P$, where

$$\Phi = (R^*C)^{\#}$$
$$P = (\nu y)\,(\overline{y}\langle x, x\rangle \mid y(z_1, z_2).\,\mathbf{read}(z_1).\mathbf{close}(z_2)).$$

Let $\Gamma = y : \mathbf{chan}\langle(z_1, z_2)z_1^R.z_2^C\rangle, x : \mathbf{res}$. Then, the following judgment holds for the output and input processes.

$$\Gamma \triangleright \overline{y}\langle x, x\rangle : \overline{y}.\, x^R.x^C$$
$$\Gamma \triangleright y(z_1, z_2).\,\mathbf{read}(z_1).\mathbf{close}(z_2) : y.\,\mathbf{0}$$

Here, we have used subtyping relations $\langle x/z_1, x/z_2\rangle z_1^R.z_2^C \approx x^R.x^C$ and $z_1^R.z_2^C{\uparrow}_{\{z_1,z_2\}} \approx \mathbf{0}$. By using (T-Par) and (T-New), we obtain

$$x : \mathbf{res} \triangleright P : (\nu y)\,(\overline{y}.\,x^R.x^C \mid y)$$

$$\Gamma \triangleright \mathbf{0} : \mathbf{0} \qquad \text{(T-Zero)}$$

$$\frac{\Gamma \triangleright P : A_2 \quad \Gamma \triangleright x : \mathbf{chan}\langle(\widetilde{y} : \widetilde{\sigma})A_1\rangle \quad \Gamma \triangleright \widetilde{v} : \widetilde{\sigma}}{\Gamma \triangleright \overline{x}\langle\widetilde{v}\rangle.\, P : \overline{x}.\, (\langle\widetilde{v}/\widetilde{y}\rangle A_1 \mid A_2)} \qquad \text{(T-Out)}$$

$$\frac{\Gamma, \widetilde{y} : \widetilde{\sigma} \triangleright P : A_2 \quad \Gamma \triangleright x : \mathbf{chan}\langle(\widetilde{y} : \widetilde{\sigma})A_1\rangle \quad A_2{\downarrow_{\{\widetilde{y}\}}} \leq A_1}{\Gamma \triangleright x(\widetilde{y}).\, P : x.\, (A_2{\uparrow_{\{\widetilde{y}\}}})} \qquad \text{(T-In)}$$

$$\frac{\Gamma \triangleright P_1 : A_1 \qquad \Gamma \triangleright P_2 : A_2}{\Gamma \triangleright P_1 \mid P_2 : A_1 \mid A_2} \qquad \text{(T-Par)}$$

$$\frac{\Gamma \triangleright P : A}{\Gamma \triangleright *P : *A} \qquad \text{(T-Rep)}$$

$$\frac{\Gamma \triangleright v : \mathbf{bool} \qquad \Gamma \triangleright P : A \qquad \Gamma \triangleright Q : A}{\Gamma \triangleright \mathbf{if}\ v\ \mathbf{then}\ P\ \mathbf{else}\ Q\ : A} \qquad \text{(T-If)}$$

$$\frac{\Gamma, x : \mathbf{chan}\langle(\widetilde{y} : \widetilde{\sigma})A_1\rangle \triangleright P : A_2}{\Gamma \triangleright (\nu x)\, P : (\nu x)\, A_2} \qquad \text{(T-New)}$$

$$\frac{\Gamma \triangleright P : A \qquad \Gamma \triangleright x : \mathbf{res}}{\Gamma \triangleright \mathbf{acc}_\xi(x).P : x^\xi.A} \qquad \text{(T-Acc)}$$

$$\frac{\Gamma, x : \mathbf{res} \triangleright P : A \qquad \mathbf{traces}_x(A) \subseteq \Phi}{\Gamma \triangleright (\mathfrak{N}^\Phi x)P : A{\uparrow_{\{x\}}}} \qquad \text{(T-NewR)}$$

$$\frac{\Gamma \triangleright P : A' \qquad A' \leq A}{\Gamma \triangleright P : A} \qquad \text{(T-Sub)}$$

Figure 4: Typing Rules

Using (T-Sub) with $(\nu y)\,(\overline{y}.\, x^R.x^C \mid y) \approx x^R.x^C$ we get

$$x : \mathbf{res} \triangleright P : x^R.x^C$$

Since $\mathbf{traces}_x(x^R.x^C)) \subseteq (R^*C)^\#$, we obtain $\emptyset \triangleright (\mathfrak{N}^\Phi x)P : \mathbf{0}$ by using (T-NewR) and (T-Sub). $\square$

**Example 3.11.** Recall Example 2.9:

$$\begin{aligned}
P &= (\nu s)\,(*s(n,x,r).\, P_1 \ \mid\ (\mathfrak{N}^\Phi x)P_2) \\
P_1 &= \mathbf{if}\ n = 0\ \mathbf{then}\ \overline{r}\langle\rangle \\
&\quad\ \mathbf{else}\ (\nu r')\,(\overline{s}\langle n-1, x, r'\rangle \mid r'().\, \mathbf{read}(x).\overline{r}\langle\rangle) \\
P_2 &= (\nu r)\,(\mathbf{init}(x).\overline{s}\langle 100, x, r\rangle \mid r().\, \mathbf{close}(x)) \\
\Phi &= (IR^*C)^\#
\end{aligned}$$

Let $A_1 = \mu\alpha.(\overline{r} \oplus (\nu r')\,(\langle r'/r\rangle\alpha | r'.\, x^R.\overline{r})$ and
let $\Gamma = s{:}\mathbf{chan}\langle(n{:}\mathbf{int},\ x{:}\mathbf{res},\ r{:}\mathbf{chan}\langle\rangle)\, A_1\rangle$. Then

$$\begin{aligned}
&\Gamma, n{:}\mathbf{int}, x{:}\mathbf{res}, r{:}\mathbf{chan}\langle\rangle \triangleright P_1 : A_1 \\
&\Gamma \triangleright *s(n,x,r).\, P_1 : *s.\, (A_1{\uparrow_{\{n,x,r\}}}) \approx *s \\
&\Gamma \triangleright P_2 : (\nu r)\,(x^I.A_1 | r.\, x^C)
\end{aligned}$$

So long as $\mathbf{traces}_x((\nu r)\,(x^I.A_1|r.\,x^C)) \subseteq \Phi$, we obtain $\emptyset \triangleright P : \mathbf{0}$. See Section 4.3 for the algorithm that establishes $\mathbf{traces}_x(\cdot) \subseteq \Phi$. □

**Remark 3.12.** The type $A_1$ in the example above demonstrates how recursion, hiding, and renaming are used together. In general, in order to type a recursive process of the form $*s(x).\,(\nu y)\,(\cdots \overline{s}\langle y\rangle \cdots)$, we need to find a type that satisfies $(\nu y)\,(\cdots \langle y/x\rangle A \cdots) \leq A$. Moreover, for the type inference (in Section 4), we must find the *least* such $A$. Thanks to the type constructors for recursion, hiding, and renaming, we can always do that: $A$ can be expressed by $\mu\alpha.(\nu y)\,(\cdots \langle y/x\rangle \alpha \cdots)$ (recall Lemma 3.8.4).

**Remark 3.13.** A reader may wonder why the rules (T-Out) and (T-In) are asymmetric, in the sense that information about the continuation of a receiver process is transferred to a sender process but not vice versa. That design choice comes from the observation that a channel or resource exchanged between a sender and a receiver are, in general, statically known only to the sender, so that we have to put information about the behavior on the channel or resource into the type of the sender. For example, consider the process $((\nu y)\,(\overline{x}\langle y\rangle \mid \cdots) \mid x(z).\,\overline{z}\langle\,\rangle$. Since the receiver $x(z).\,\overline{z}\langle\,\rangle$ is not in the scope of $y$, we have to put the information that $y$ will be used for output into the type of the sender $\overline{x}\langle y\rangle$ (as $\overline{x}.\,\overline{y}$). It is still useful and possible to recover the symmetry in the treatment of senders and receivers to some extent: see Section 8 of our previous paper [10].

The following theorem states that no well-typed process performs an invalid access to a resource.

**Theorem 3.14** (type soundness (safety))**.** *Suppose that $P$ is safe. If $\Gamma \triangleright P : A$ and $P \longrightarrow^* Q$, then $Q$ is safe.*

*Proof.* We make use of the following lemma:

- **Subject-reduction.** If $P \xrightarrow{L} P'$ and $\Gamma \triangleright P : A$ then $A \xRightarrow{L} A'$ and $\Gamma \triangleright P' : A'$. Proof: see Appendix B.

For the proof of the theorem, we focus on just a single reduction step. By the Lemma we know that judgements are preserved by reduction; we must show that safety is also preserved, by induction on the derivation of reduction. The only interesting case is (R-NewR1), $(\mathfrak{N}^{\Phi}x)P \xrightarrow{\tau} (\mathfrak{N}^{\Phi^{-\xi}}x)P'$, since the other rules do not alter trace-sets $\Phi$. In this case, we are given $\Gamma \triangleright P : A$, $\mathbf{traces}_x(A) \subseteq \Phi$, and $P \xrightarrow{x^\xi} P'$. By the Lemma, $A \xRightarrow{x^\xi} A'$ for some $\Gamma \triangleright P' : A'$. Assume $(\mathfrak{N}^{\Phi}x)P$ is safe; hence so is $P$; by the induction hypothesis so is $P'$. From the conditions $\mathbf{traces}_x(A) \subseteq \Phi$ and $A \xRightarrow{x^\xi} A'$, we get $\xi \in \mathbf{traces}_x(A) \subseteq \Phi$, so that $\epsilon \in \Phi^{-\xi} \neq \emptyset$. So, $(\mathfrak{N}^{\Phi^{-\xi}}x)P'$ is safe. □

## 4. Type Inference Algorithm

This section discusses an algorithm which takes a closed process $P$ as an input and checks whether $\emptyset \triangleright P : \mathbf{0}$ holds. As in similar type systems [11, 15], the algorithm consists of the following steps.

(1) Extract constraints on type variables based on the (syntax-directed version of) typing rules.
(2) Reduce constraints to trace inclusion constraints of the form $\{\mathbf{traces}_{x_1}(A_1) \subseteq \Phi_1, \ldots, \mathbf{traces}_{x_n}(A_n) \subseteq \Phi_n\}$

$$\emptyset \triangleright_{sd} \mathbf{0} : \mathbf{0} \qquad \text{(T-SD-Zero)}$$

$$\frac{\Gamma_0 \triangleright_{sd} P : A_2 \qquad \Gamma_i \triangleright v_i : \sigma_i \ \text{(for each } i \in \{1,\dots,n\}\text{)}}{\Gamma_0 \cup \widetilde{\Gamma} \cup (x : \mathbf{chan}\langle(\widetilde{y} : \widetilde{\sigma})A_1\rangle) \triangleright_{sd} \overline{x}\langle\widetilde{v}\rangle.\, P : \overline{x}.\,(\langle\widetilde{v}/\widetilde{y}\rangle A_1 \mid A_2)} \qquad \text{(T-SD-Out)}$$

$$\frac{\Gamma \triangleright_{sd} P : A_2 \qquad A_2{\downarrow_{\{\widetilde{y}\}}} \leq A_1 \qquad wd(\Gamma \cup \widetilde{y} : \widetilde{\sigma})}{(\Gamma \backslash \{\widetilde{y}\}) \cup x : \mathbf{chan}\langle(\widetilde{y} : \widetilde{\sigma})A_1\rangle \triangleright_{sd} x(\widetilde{y}).\, P : x.\, A_2{\uparrow_{\{\widetilde{y}\}}}} \qquad \text{(T-SD-In)}$$

$$\frac{\Gamma_1 \triangleright_{sd} P_1 : A_1 \qquad \Gamma_2 \triangleright_{sd} P_2 : A_2}{\Gamma_1 \cup \Gamma_2 \triangleright_{sd} P_1 \mid P_2 : A_1 \mid A_2} \qquad \text{(T-SD-Par)}$$

$$\frac{\Gamma \triangleright_{sd} P : A}{\Gamma \triangleright_{sd} *P : *A} \qquad \text{(T-SD-Rep)}$$

$$\frac{\Gamma_0 \triangleright v : \mathbf{bool} \qquad \Gamma_1 \triangleright_{sd} P : A_1 \qquad \Gamma_2 \triangleright_{sd} Q : A_2 \qquad A_1 \leq A \qquad A_2 \leq A}{\Gamma_0 \cup \Gamma_1 \cup \Gamma_2 \triangleright_{sd} \mathbf{if}\ v\ \mathbf{then}\ P\ \mathbf{else}\ Q : A} \qquad \text{(T-SD-If)}$$

$$\frac{\Gamma \triangleright_{sd} P : A_2 \qquad wd(\Gamma \cup (x : \mathbf{chan}\langle(\widetilde{x} : \widetilde{\tau})A_1\rangle))}{\Gamma \backslash \{x\} \triangleright_{sd} (\nu x)\, P : (\nu x)\, A_2} \qquad \text{(T-SD-New)}$$

$$\frac{\Gamma \triangleright_{sd} P : A}{\Gamma \cup (x : \mathbf{res}) \triangleright_{sd} \mathbf{acc}_\xi(x).P : x^\xi.A} \qquad \text{(T-SD-Acc)}$$

$$\frac{\Gamma \triangleright_{sd} P : A \qquad \mathbf{traces}_x(A) \subseteq \Phi \qquad wd(\Gamma \cup (x : \mathbf{res}))}{\Gamma \backslash \{x\} \triangleright_{sd} (\mathfrak{N}^\Phi x)P : A'{\uparrow_{\{x\}}}} \qquad \text{(T-SD-NewR)}$$

Figure 5: Syntax Directed Typing Rules

(3) Decide whether the constraints are satisfied.

The algorithm for Step 3 is sound but not complete.

We give an overview of each step below. The first two steps are almost the same as those in the previous work.

### 4.1. Step 1: Extracting Constraints.

The typing rules presented in Section 3 can be transformed to the syntax-directed typing rules shown in Figure 5. In the figure, $\Gamma_1 \cup \Gamma_2$ is the type environment obtained by merging both bindings, and defined only if $\Gamma_1(x) = \Gamma_2(x)$ for every $x \in dom(\Gamma_1) \cap dom(\Gamma_2)$. Type equality here is syntactic equality up to $\alpha$-renaming. And $wd(\Gamma_1 \cup \Gamma_2)$ means that $\Gamma_1 \cup \Gamma_2$ is well-defined. The two sets of typing rules are equivalent in the following sense: If $\Gamma \triangleright P : A$ is derivable, then there exists $A'$ such that $A' \leq A$ holds and $\Gamma \triangleright_{sd} P : A'$ is derivable. Conversely, if $\Gamma \triangleright_{sd} P : A$ is derivable, so is $\Gamma \triangleright P : A$.

Based on the syntax-directed rules, we obtain the algorithm in Figure 6, which takes a process $P$ and outputs a triple consisting of a type environment $\Gamma$, a behavioral type $A$, and a set $C$ of constraints. In Figure 6, $\Gamma_1 \otimes \cdots \otimes \Gamma_n$ is defined to be $(\Gamma, C)$ where $\Gamma$ and $C$ are given by:

$$\begin{array}{l} dom(\Gamma) = dom(\Gamma_1) \cup \cdots \cup dom(\Gamma_n) \\ \Gamma(x) = \Gamma_i(x) \text{ where } x \in dom(\Gamma_i) \backslash (dom(\Gamma_1) \cup \cdots \cup dom(\Gamma_{i-1})) \\ C = \{\Gamma_i(x) = \Gamma_j(x) \mid x \in dom(\Gamma_i) \cap dom(\Gamma_j)\} \end{array}$$

The triple $(\Gamma, A, C)$ output by $PT$ satisfies the following properties:

- $\theta\Gamma \triangleright P : \theta A$ holds for any substitution $\theta$ such that $\models \theta C$.

- If $\Gamma' \triangleright P : A'$, then there exists a substitution $\theta$ such that $\theta\Gamma \subseteq \Gamma'$ and $\theta A \leq A'$.

Here, $\Gamma$ and $A$ may contain variables representing unknown behavioral types and value types. $C$ is a set of constraints on them, and the substitution $\theta$ above replaces them with closed behavioral types and value types. Intuitively, the triple $(\Gamma, A, C$ expresses a set of type judgments for $P$. The first property above says that the triple contains only valid judgments, while the second property says that every valid judgment is subsumed by the triple.

We do not give a formal proof of the above properties; As usual, they can be proved by induction on the structure of $P$.

4.2. **Step 2: Reducing Constraints.** Given a closed process $P$, $PT(P)$ produces a triple $(\emptyset, A, C)$. The set $C$ of constraints consists of unification constraints on value types (where all the behavioral types occurring in them are variables), constraints of the form $\mathbf{isChan}(\sigma)$ (which means that $\sigma$ is a channel type), subtype constraints on behavioral types of the form $\alpha \geq A$, and constraints of the form $\mathbf{traces}_x(A) \subseteq \Phi$. We can remove the first two kinds of constraints (unification constraints on value types and $\mathbf{isChan}(\sigma)$) by applying the standard unification algorithm. Thus, we obtain the following constraints:

$$\begin{array}{l}\{\alpha_1 \geq A_1, \ldots, \alpha_n \geq A_n,\\ \mathbf{traces}_{x_1}(B_1) \subseteq \Phi_1, \ldots, \mathbf{traces}_{x_m}(B_m) \subseteq \Phi_m\}\end{array}$$

Here, we can assume that $\alpha_1, \ldots, \alpha_n$ are different from each other, since $\alpha \geq A_1$ and $\alpha \geq A_2$ can be replaced with $\alpha \geq A_1 \oplus A_2$ by Lemma 3.8. We can also assume that $\{\alpha_1, \ldots, \alpha_n\}$ contains all the type variables in the constraint, since otherwise we can always add the tautology $\alpha \geq \alpha$. Each subtype constraint $\alpha \geq A$ can also be replaced by $\alpha \geq \mu\alpha.A$, by using Lemma 3.8. Therefore, the above constraints can be further reduced, by Lemma 3.8, to:

$$\{\mathbf{traces}_{x_1}([\widetilde{A'}/\widetilde{\alpha}]B_1) \subseteq \Phi_1, \ldots, \mathbf{traces}_{x_m}([\widetilde{A'}/\widetilde{\alpha}]B_m) \subseteq \Phi_m\}$$

Here, $A'_1, \ldots, A'_n$ are the least solutions for the subtype constraints.

Thus, we have reduced type checking to the validity of trace inclusion constraints of the form $\mathbf{traces}_x(A) \subseteq \Phi$.

**Example 4.1.** Recall Example 2.9. By applying the algorithm $PT$ and the first part of Step 2, we obtain the following constraints:

$$\begin{array}{l}\mathbf{traces}_x((\nu r)\,(x^I.\overline{s}.\,\alpha_1 \,|\, r.\,x^C)) \subseteq (IR^*C)^\#\\ \alpha_1 \geq \overline{r}.\,\alpha_2 \oplus (\nu r')\,(\overline{s}.\,\langle r'/r\rangle\alpha_1 \,|\, r'.\,x^R.\overline{r}.\,\alpha_2)\downarrow_{\{n,x,r\}}\\ \alpha_2 \geq \alpha_2\end{array}$$

By applying the second part of Step 2, we obtain $\mathbf{traces}_x(A_1) \subseteq (IR^*C)^\#$ where

$$\begin{array}{rcl}A_1 &=& (\nu r)\,(x^I.\overline{s}.\,A_2 \,|\, r.\,x^C)\\ A_2 &=& \mu\alpha_1.\overline{r}.\,A_3 \oplus (\nu r')\,(\overline{s}.\,\langle r'/r\rangle\alpha_1 \,|\, r'.\,x^R.\overline{r}.\,A_3)\downarrow_{\{n,x,r\}}\\ A_3 &=& \mu\alpha_2.\alpha_2.\end{array}$$

$PTv(x) = (x : \rho, \rho)$ (where $\rho$ fresh)
$PTv(b) = (\emptyset, \mathbf{bool})$ if $b \in \{\mathbf{true}, \mathbf{false}\}$

$PT(\mathbf{0}) = (\emptyset, \mathbf{0}, \emptyset)$
$PT(\overline{x}\langle\widetilde{v}\rangle.\, P_0) =$
  **let** $(\Gamma_i, \sigma_i) = PTv(v_i)$
    $(\Gamma_0, A_0, C_0) = PT(P_0)$
    $(\Gamma, C) = \Gamma_0 \otimes (x : \mathbf{chan}\langle(\widetilde{y} : \widetilde{\sigma})\alpha\rangle) \otimes \Gamma_1 \otimes \cdots \otimes \Gamma_n$
  **in** $(\Gamma, \overline{x}.\, ([\widetilde{v}/\widetilde{y}]\alpha \mid A_0), C)$ (where $\alpha$ fresh)
$PT(x(\widetilde{y}).\, P_0) =$
  **let** $(\Gamma_0, A_0, C_0) = PT(P_0)$
    $(\Gamma_1, C_1) = \Gamma_0 \otimes (x : \mathbf{chan}\langle(\widetilde{y} : \widetilde{\rho})\alpha\rangle) \otimes (\widetilde{y} : \widetilde{\rho})$
  **in** $(\Gamma \backslash \widetilde{y}, x.\, A_0 \uparrow_{\{\widetilde{y}\}}, C_0 \cup C_1 \cup \{\alpha \geq A_0 \downarrow_{\{\widetilde{y}\}}\})$
    (where $\alpha$, $\widetilde{\rho}$ fresh)
$PT(P_0 \mid P_1) =$
  **let** $(\Gamma_0, A_0, C_0) = PT(P_0)$
    $(\Gamma_1, A_1, C_1) = PT(P_1)$
    $(\Gamma_2, C_2) = \Gamma_0 \otimes \Gamma_1$
  **in** $(\Gamma_2, A_0 \mid A_1, C_0 \cup C_1 \cup C_2)$
$PT(\mathbf{if}\ v\ \mathbf{then}\ P_0\ \mathbf{else}\ P_1\ ) =$
  **let** $(\Gamma_0, A_0, C_0) = PT(P_0)$
    $(\Gamma_1, A_1, C_1) = PT(P_1)$
    $(\Gamma_2, \sigma) = PTv(v)$
    $(\Gamma, C_2) = \Gamma_0 \otimes \Gamma_1 \otimes \Gamma_2$
  **in** $(\Gamma, A_0 \oplus A_1, C_0 \cup C_1 \cup C_2 \cup \{\sigma = \mathbf{bool}\})$
$PT((\nu x)\, P_0) =$
  **let** $(\Gamma_0, A_0, C_0) = PT(P_0)$
    $C_1 = \mathbf{if}\ x \in dom(\Gamma_0) \mathbf{then}\ \{\mathbf{isChan}(\Gamma_0(x))\} \mathbf{else}\ \emptyset$
  **in** $(\Gamma_0 \backslash \{x\}, (\nu x)\, A_0, C_0 \cup C_1)$
$PT(*P_0) =$
  **let** $(\Gamma_0, A_0, C_0) = PT(P_0)$
  **in** $(\Gamma_0, *A_0, C_0)$
$PT(\mathbf{acc}_\xi(x).P_0) =$
  **let** $(\Gamma_0, A_0, C_0) = PT(P_0)$
    $(\Gamma_1, C_1) = \Gamma_0 \otimes (x : \mathbf{res})$
  **in** $(\Gamma_1, x^\xi.A_0, C_0 \cup C_1)$
$PT((\mathfrak{N}^\Phi x)P_0) =$
  **let** $(\Gamma_0, A_0, C_0) = PT(P_0)$
    $(\Gamma_1, C_1) = \Gamma_0 \otimes (x : \mathbf{res})$
  **in** $(\Gamma_1 \backslash \{x\}, A_0 \uparrow_{\{x\}}, C_0 \cup C_1 \cup \{\mathbf{traces}_x(A_0) \subseteq \Phi\})$

Figure 6: A Type Inference Algorithm

4.3. **Step 3: Constraint Solving.** We present an approximation algorithm for checking a trace inclusion constraint $\mathbf{traces}_x(A) \subseteq \Phi$ when the trace set $\Phi$ is a regular language. (Actually, we can extend the algorithm to deal with the case where $\Phi$ is a deterministic Petri net language: see Remark 4.6.)

We first describe the algorithm with an example. In Example 4.1 above, we have reduced the typability of the process to the equivalent constraint $\mathbf{traces}_x(A_1) \subseteq \Phi$ where

$\Phi = (IR^*C)^\#$ and

$$A_1\downarrow_{\{x\}} \approx (\nu r)\,(x^I.A_2'' \mid r.\,x^C)$$
$$A_2'' = \overline{r} \oplus (\nu r')\,(\langle r'/r\rangle A_2'' \mid r'.\,x^R.\overline{r})$$

Here, we have removed $A_3 = \mu\alpha.\alpha$ since $A_3 \approx \mathbf{0}$.

**Step 3-1.** Approximate the behavior of $A_1\downarrow_{\{x\}}$ by a Petri net [22] $N_{A_1,x}$. This part is similar to the translation of usage expressions into Petri nets in Kobayashi's previous work [16, 15, 12]. Since the behavioral types are more expressive (having recursion, hiding, and renaming), however, we need to approximate the behavior of a behavioral type unlike in the previous work. In this case $A_1\downarrow_{\{x\}}$ is infinite. To make it tractable we make a sound approximation $A_1'$ by pushing $(\nu)$ to top level, and we eliminate $\langle r'/r\rangle$:

$$A_1' = (\nu r, r')\,(x^I.A_2' \mid r.\,x^C)$$
$$A_2' = \overline{r} \oplus (A_3' \mid r'.\,x^R.\overline{r})$$
$$A_3' = \overline{r'} \oplus (A_3' \mid r'.\,x^R.\overline{r'})$$

Then $N_{A_1',x}$ is as pictured. (Here we treat $A_1 \oplus A_2$ as $\tau.A_1 \oplus \tau.A_2$ for clarity. We also use a version of Petri nets with labeled transitions.)

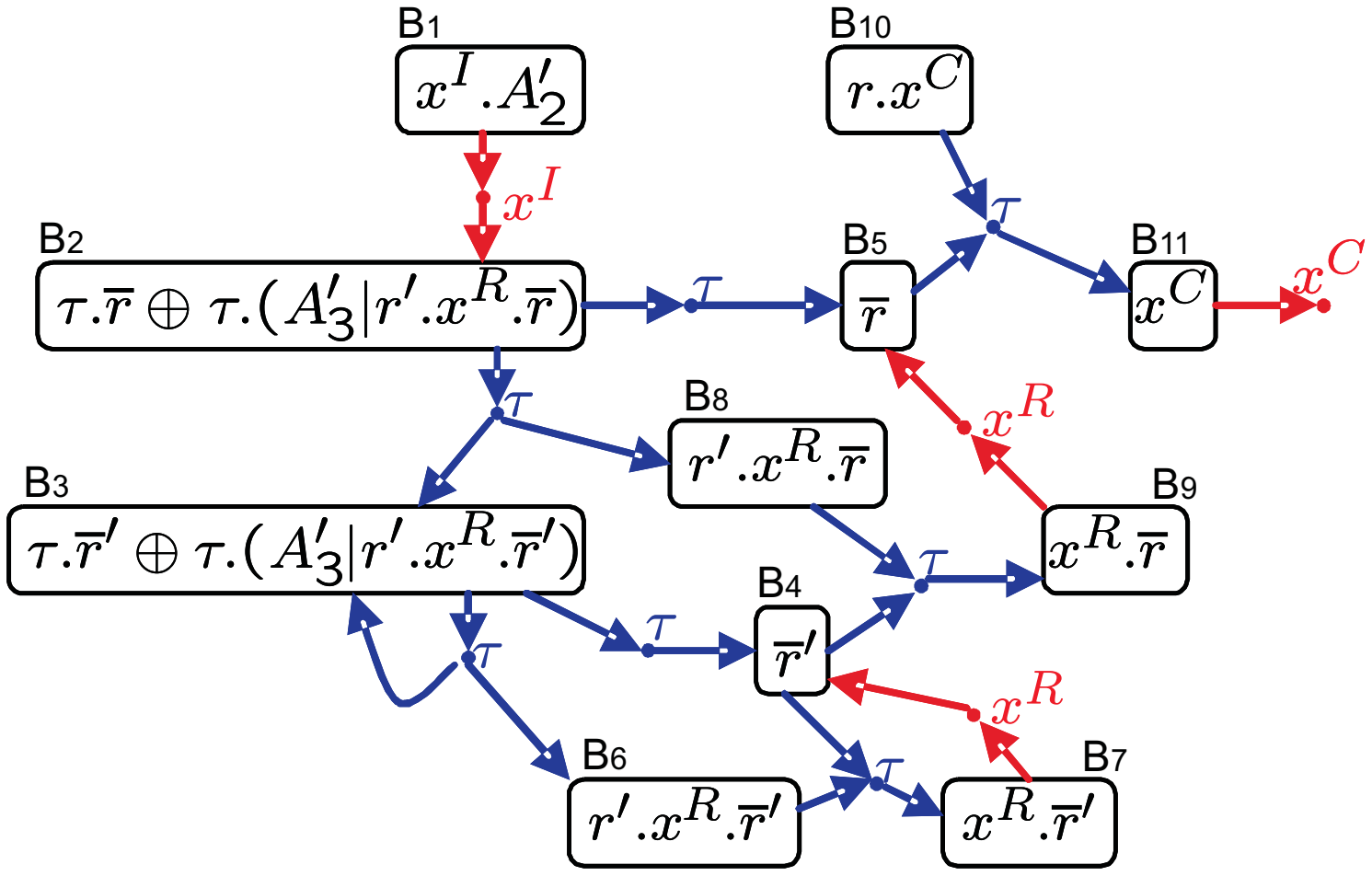

The rectangles are the places of the net, and the dots labeled by $\tau, x^R$, etc. are the transitions of the net. Write $i_x$ for the number of tokens at node $B_x$. The behavior $A_1'$ corresponds to the initial marking $\{i_1{=}1,\ i_{10}{=}1\}$. We say that the nodes $\widetilde{B}$ together with the restricted names $(r, r')$ constitute a *basis* for $A_1'$. Note here that $\mathbf{traces}_x(A_1) \subseteq \mathbf{traces}_x(A_1') = \mathbf{ptraces}(N_{A_1',x})$ where $\mathbf{ptraces}(N_{A_1',x})$ is the set of traces of the Petri net. Thus, $\mathbf{ptraces}(N_{A_1',x}) \subseteq \Phi$ is a sufficient condition for $\mathbf{traces}_x(A_1) \subseteq \Phi$ . The key point here is that $A_1'$ still has infinite states, but all its reachable states can be expressed in the form $(\nu r, r')\,(i_1 B_1 \mid \cdots \mid i_{11} B_{11})$ (where $i_k B_k$ is the parallel composition of $i_k$ copies of $B_k$), a linear combination of finitely many processes $\widetilde{B}$. That is why we could express $A_1'$ by the Petri net as above.

**Step 3-2.** Construct a deterministic, minimized automaton $M_\Phi$ that accepts the language $\Phi$. Here the initial marking is $\{i_{12}{=}1\}$.

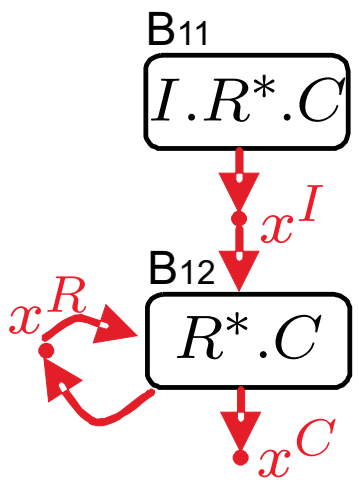

**Step 3-3.** Construct another Petri net $N_{A'_1,x} \parallel M_\Phi$ from $N_{A'_1,x}$ and $M_\Phi$, which simulates the behavior of $P_A$ and $M_\Phi$ simultaneously, so that the problem of $\mathbf{traces}_x(A'_1)(= \mathbf{ptraces}(N_{A'_1,x})) \subseteq \Phi$ is equivalent to a reachability problem of $N_{A'_1,x} \parallel M_\Phi$. In the example, $N_{A'_1,x} \parallel M_\Phi$ has the initial marking $\{i_1{=}1, i_{10}{=}1, i_{12}{=}1\}$ and transitions such as $B_1|B_{12} \xrightarrow{I} B_2|B_{13}$. $\mathbf{ptraces}(N_{A'_1,x}) \subseteq \Phi$ if and only if the following unsafe state is unreachable.

$$(i_1{>}0 \wedge i_{12}{=}0) \vee (i_7{>}0 \wedge i_{13}{=}0) \vee (i_9{>}0 \wedge i_{13}{=}0) \vee (i_{11}{>}0 \wedge i_{13}{=}0)$$

To explain, if $i_1 > 0 \wedge i_{12}{=}0$ then the behavior is able to make an $R$ transition but the specification automaton $M_\Phi$ is not able.

**Step 3-4.** Use an approximation algorithm to decide the reachability problem of $N_{A'_1,x} \parallel M_\Phi$, in a manner similar to Kobayashi's type-based analyzer `TyPiCal` [12] for the $\pi$-calculus.

The above steps 3-1, 3-2, and 3-3 are described in more detail below. See Section 6 for Step 3-4.

4.3.1. *Step 3-1: Construction of* $N_{A,x}$. We first introduce the notion of a *basis*. The basis is analogous to that of a vector space; Each state is expressed as a linear combination of elements of the basis.

**Definition 4.2.** A pair $(\{y_1,\ldots,y_m\},\{B_1,\ldots,B_n\})$ is a *basis* of $A$ if all of the following conditions are satisfied:

- $A \approx (\nu y_1)\cdots(\nu y_m)\,(i_1 B_1 \,|\, \cdots \,|\, i_n B_n)$ for some $i_1,\ldots,i_n \in \mathbf{Nat}$.
- If $B_j \xrightarrow{l} C$, then there exist $i_1,\ldots,i_n \in \mathbf{Nat}$ such that $C \approx i_1 B_1 \,|\, \cdots \,|\, i_n B_n$.
- For each $B_j$, there are only finitely many $C$ (up to $\approx$) such that $B_j \xrightarrow{l} C$.

Note that if $(\{\widetilde{y}\},\{B_1,\ldots,B_n\})$ is a basis of $A$, then whenever $A \Longrightarrow A'$, there exist $i_1,\ldots,i_n$ such that $A' \approx (\nu\widetilde{y})\,(i_1 B_1 \,|\, \cdots \,|\, i_n B_n)$. Let us write *Index*$(C)$ for $(i_1,\ldots,i_n)$ such that $C \approx i_1 B_1 \,|\, \cdots i_n B_n$. (If there are more than one such tuple, *Index*$(C)$ picks one among them.) Therefore, if $A{\downarrow_{\{x\}}}$ has a basis, the behavior of $A{\downarrow_{\{x\}}}$ is simulated by the (labeled) Petri net $N_{A,x,(\{\widetilde{y}\},\{\widetilde{B}\})}$ given below. Here, we use a process-like syntax to represent the elements of a Petri net rather than the standard tuple notation $(P,T,F,W,M_0)$. A marking state $m$ which has $i_k$ tokens for each place $p_k$ $(k \in \{1,\ldots,n\})$ is written $i_1 p_1 \,|\, \cdots \,|\, i_n p_n$. A transition that consumes a marking $m_1$ and produces $m_2$ is expressed by $m_1 \xrightarrow{\gamma} m_2$, where $\gamma$ is the label of the transition.

- The set $P$ of places is $\{p_{B_1},\ldots,p_{B_n}\}$.

- The initial marking $m_I$ is $i_1 p_{B_1} \,|\, \cdots \,|\, i_n p_{B_n}$ where $A{\downarrow}_{\{x\}} \approx (\nu\widetilde{y})\,(i_1 B_1 \,|\, \cdots \,|\, i_n B_n)$.
- The set of transitions consists of:
  - $p_{B_j} \xrightarrow{\tau} i_1 p_{B_1} \,|\, \cdots \,|\, i_n p_{B_n}$ where $\mathit{Index}(C) = (i_1, \ldots, i_n)$, for each $B_j \xrightarrow{\tau} C$.
  - $p_{B_j} \xrightarrow{\xi} i_1 p_{B_1} \,|\, \cdots \,|\, i_n p_{B_n}$ where $\mathit{Index}(C) = (i_1, \ldots, i_n)$, for each $B_j \xrightarrow{x^\xi} C$.
  - $p_{B_j} \,|\, p_{B_{j'}} \xrightarrow{\tau} (i_1 + i'_1) p_{B_1} \,|\, \cdots \,|\, (i_n + i'_n) p_{B_n}$ where $\mathit{Index}(C) = (i_1, \ldots, i_n)$ and $\mathit{Index}(C') = (i'_1, \ldots, i'_n)$, for each pair of transitions $B_j \xrightarrow{\overline{z}} C$ and $B_{j'} \xrightarrow{z} C'$ such that $z \in \{\widetilde{y}\}$.

From now on we omit the basis and just write $N_{A,x}$ for $N_{A,x,(\{\widetilde{y}\},\{\widetilde{B}\})}$. Let us write $\mathbf{ptraces}(N_{A,x})$ for the set:

$$\{\xi_1 \cdots \xi_k \mid m_I \stackrel{\xi_1}{\Longrightarrow} \cdots \stackrel{\xi_k}{\Longrightarrow} m'\}$$

where $\stackrel{\xi}{\Longrightarrow}$ means $\xrightarrow{\tau}^{*} \xrightarrow{\xi} \xrightarrow{\tau}^{*}$. By the construction of $N_{A,x}$, $\mathbf{ptraces}(N_{A,x}) = \mathbf{traces}_x(A)$.

The construction of $N_{A,x}$ outlined above can be applied only when a basis of $A{\downarrow}_x$ can be found (by some heuristic algorithm). If $A{\downarrow}_x$ has no basis or cannot be found, we approximate $A{\downarrow}_x$ by moving all the $\nu$-prefixes to the top-level; for example, $y.\,(\nu x)\,A$, $*(\nu x)\,A$ and $\mu\alpha.(\nu x)\,A$ are replaced by $(\nu x)\,(y.\,A)$, $(\nu x)\,{*A}$, and $(\nu x)\,\mu\alpha.A$ respectively. Let $A'$ be the approximation of $A{\downarrow}_{\{x\}}$. It is easy to prove that $A'$ is a sound approximation of $A{\downarrow}_{\{x\}}$, in the sense that $\mathbf{traces}_x(A) \subseteq \mathbf{traces}_x(A')$.

We can compute a basis of $A'$ as follows (see Appendix D for more details). Since $\nu$-prefixes do not appear inside recursion, we can first eliminate the constructors $\cdot{\uparrow}_S$, $\cdot{\downarrow}_S$, and $\langle\widetilde{y}/\widetilde{x}\rangle$. Let $(\nu\widetilde{y})\,A''$ be the resulting expression, where $A''$ does not contain $\cdot{\uparrow}_S$, $\langle\widetilde{y}/\widetilde{x}\rangle$, or $(\nu x)$. Let $\mathbf{B}$ be the set of behavioral types that are subexpressions of the behavioral types obtained from $A''$ by expanding recursive types and do not contain "unnecessary" unfolding $[\mu\alpha.A/\alpha]A$. Then, $\mathbf{B}$ is a finite set, and $(\{\widetilde{y}\}, \mathbf{B})$ is a basis of $A'$. We can therefore construct a Petri net $N_{A',x}$. By the construction, $\mathbf{ptraces}(N_{A',x}) = \mathbf{traces}_x(A') \supseteq \mathbf{traces}_x(A)$, so that $\mathbf{ptraces}(N_{A',x}) \subseteq \Phi$ is a sufficient condition for $\mathbf{traces}_x(A) \subseteq \Phi$.

4.3.2. *Steps 3-2 and 3-3: Construction of $N_{A,x} \,\|\, M_\Phi$ and reduction of $\mathbf{traces}_x(A)$ to a reachability problem.* Let $P_{N_{A,x}}$ and $T_{N_{A,x}}$ be the sets of places and transitions of $N_{A,x}$ respectively. Let $M_\Phi$ be a minimized deterministic automaton[4] that accepts $\Phi$, and let $Q_\Phi$ be its set of states and $\delta_\Phi$ be its transition function.

**Definition 4.3.** The *composition* of $N_{A,x}$ and $M_\Phi$, written $N_{A,x} \,\|\, M_\Phi$, is defined as follows:

- The set of places is $P_{N_{A,x}} \cup Q_\Phi$
- The set of transitions is:

$$\begin{array}{l}\{(m|q) \xrightarrow{\xi} (m'|q') \mid (m \xrightarrow{\xi} m') \in T_{N_{A,x}} \wedge \delta_\Phi(q,\xi) = q'\} \\ \cup\{m \xrightarrow{\tau} m' \mid (m \xrightarrow{\tau} m') \in T_{N_{A,x}}\}\end{array}$$

- Initial state is $m_I \,|\, q_I$ where $m_I$ is the initial state of $N_{A,x}$ and $q_I$ is the initial state of $M_\Phi$.

[4]Note that since $\Phi$ is prefix-closed, all the states of the minimized automaton are accepting states.

Now, $\mathbf{ptraces}(N_{A,x}) \subseteq \Phi$ can be reduced to the reachability problems of $N_{A,x} \parallel M_\Phi$.

**Theorem 4.4.** *$\mathbf{ptraces}(N_{A,x}) \subseteq \Phi$ if and only if no marking $m \,|\, q$ that satisfies the following conditions is reachable:*

- *$m \xrightarrow{\xi} m'$ for some $m'$ and $\xi$ in $N_{A,x}$.*
- *$\delta_\Phi(q, \xi)$ is undefined.*

Thus, we can reduce $\mathbf{ptraces}(N_{A,x}) \subseteq \Phi$ to a finite set of reachability problems of $N_{A,x} \parallel M_\Phi$. Hence $\mathbf{ptraces}(N_{A,x}) \subseteq \Phi$ is decidable [18].

**Corollary 4.5.** *$\mathbf{ptraces}(N_{A,x}) \subseteq \Phi$ if and only if for every transition rule of the form $m_1 \xrightarrow{\xi} m_2$ of $N_{A,x}$ and $q$ such that $\delta_\Phi(q, \xi)$ is undefined, no marking $m$ such that $m \geq m_1 \,|\, q$ is reachable by $N_{A,x} \parallel M_\Phi$.*

**Remark 4.6.** We can actually extend the above algorithm for checking $\mathbf{traces}_x(A) \subseteq \Phi$ to deal with the case where $\Phi$ belongs to the class of deterministic Petri net languages (more precisely, the class of P-type languages of $\lambda$-free, deterministic Petri nets [22, 21]). If $\Phi$ is the P-type language of a $\lambda$-free, deterministic Petri net, then its complement $\overline{\Phi}$ is a Petri net language [21]. Therefore, we can construct a Petri net that accepts the intersection of the language of $N_{A,x}$ and $\overline{\Phi}$ [22]), so that $\mathbf{ptraces}(N_{A,x}) \subseteq \Phi$ can be reduced to the emptiness problem of the Petri net, which is decidable due to the decidability of the reachability problem.

Some of the useful resource usage specifications are not regular languages but are deterministic Petri net language. For example, consider a stack-like resource on which, at any point of program execution, the number of times the operation *pop* has been performed is less than the number of times *push* has been performed. Such specification is expressible as a deterministic Petri net language.

## 5. Extensions

The type system given so far guarantees that no invalid resource access is performed, but not that any necessary access is performed eventually; for example, the type system does *not* guarantee that a file is eventually closed. We discuss extensions of the type system to guarantee such properties.

We are interested in type systems that satisfy either *partial liveness*[5] or the stronger *liveness* property:

- partial liveness: If $P \longrightarrow^* Q$ and $Q \not\longrightarrow$, then $Q$ does not contain any resource to which some access *must* be performed.
- liveness: In any fair reduction sequence $P \longrightarrow P_1 \longrightarrow P_2 \longrightarrow \cdots$, $P$ eventually performs all the necessary resource access. (Here, a reduction sequence is fair if an input or output action that is infinitely enabled will eventually succeed. Without the fairness assumption, no process can satisfy the liveness property in the presence of a divergent process $(\nu x)\,(\overline{x}\langle\,\rangle \,|\, {*x(\,).\,\overline{x}\langle\,\rangle}$, which is too restrictive.)

[5]This is not a standard term; actually, the partial liveness here can be viewed as the safety property that no 'bad' state is reachable such that the necessary accesses have not yet been performed but the system cannot make any move.

Our idea is to take the resource type system from the previous sections, and combine it with some existing system that annotates those communications that eventually succeed. Specifically, this existing system might be (1) deadlock-freedom [16, 15], which guarantees that the annotated communications eventually succeed unless the process diverges; the combination would then guarantee partial liveness. Or the existing system could be (2) lock-freedom [14, 15], which guarantees that the annotated communications eventually succeed even in the presence of divergence (assuming a strongly fair scheduler); the combination would then guarantee full liveness.

To formally state which resource access *must* be performed, we extend the trace sets.

**Definition 5.1.** An *extended trace set* is a set of sequences of access labels, possibly ending with a special label $\downarrow$, that is closed under the prefix operation.

Intuitively, the special label $\downarrow$ means that no further resource access need to be performed. For example, the trace set $(\{C\downarrow, RC\downarrow\})^{\#}$ means that the close operation needs to be performed, while $(\{\downarrow, R\downarrow, C\downarrow, RC\downarrow\})^{\#}$ means that the close operation need not be performed.

Now we can state the partial liveness property more formally. We write $(\widetilde{\nu \mathfrak{N}})$ for a (possibly empty) sequence of $\nu$- and $\mathfrak{N}$-binders.

**Definition 5.2.** A process $P$ is *partially live* if $\downarrow \in \Phi$ whenever $P \longrightarrow^* \preceq (\widetilde{\nu \mathfrak{N}})(\mathfrak{N}^{\Phi} x)Q \not\longrightarrow$.

5.1. **A Type System for the Partial Liveness Property.** We extend the syntax of processes to allow each input and output prefix to be annotated with information about whether the communication is guaranteed to succeed.

**Definition 5.3** ((extended) processes)**.** The set of (extended) processes is given by:

$$\begin{array}{lll} t \text{ (attributes)} & ::= & \mathbf{c} \mid \emptyset \\ P & ::= & \overline{x}_t\langle y_1, \ldots, y_n\rangle.\, P \mid x_t(y_1, \ldots, y_n).\, P \mid \cdots \end{array}$$

The attribute $\mathbf{c}$ indicates that when the annotated input or output operation appears at the top-level, the operation will succeed unless the whole process diverges, while $\emptyset$ does not give such a guarantee. We often omit tag $\emptyset$.

We assume that there exists a type system guaranteeing that any well-typed process is *well-annotated* in the sense of Definition 5.4 below. There are indeed such type systems [13, 16, 15]. Moreover, the static analysis tool `TyPiCal` [12] can automatically infer the annotations.

**Definition 5.4.** $P$ is *active*, written $\mathit{active}(P)$, if $P \preceq (\widetilde{\nu \mathfrak{N}})(\overline{x}_{\mathbf{c}}\langle \widetilde{v}\rangle.\, Q \mid R)$ or $P \preceq (\widetilde{\nu \mathfrak{N}})(x_{\mathbf{c}}(\widetilde{y}).\, Q \mid R)$. Additionally, $P$ is *well-annotated*, written *well_annotated*$(P)$, if for any $P'$ such that $P \longrightarrow^* P'$ and $\mathit{active}(P')$, there exists $P''$ such that $P' \longrightarrow P''$.

For example, $\overline{x}_{\mathbf{c}}\langle\,\rangle.\, \mathbf{0} \mid x_{\mathbf{c}}(\,).\, \overline{y}_{\emptyset}\langle\,\rangle.\, \mathbf{0}$ is well-annotated, but $\overline{x}_{\mathbf{c}}\langle\,\rangle.\, \mathbf{0} \mid x_{\mathbf{c}}(\,).\, \overline{y}_{\mathbf{c}}\langle\,\rangle.\, \mathbf{0}$ is not. Note that $x_{\emptyset}(\,).\, \overline{x}_{\mathbf{c}}\langle\,\rangle.\, \mathbf{0}$ *is* well-annotated since, although the output never succeeds, it does not appear at the top-level.

Now we introduce the type system that guarantees the partial liveness. We extend the behavioral types by extending each input, output, or $\tau$-action with an attribute to indicate whether the action is guaranteed to succeed.

$$A \;\; ::= \;\; \overline{x}_t.\, A \mid x_t.\, A \mid \tau_t.A \mid \cdots$$

| | | |
|---|---|---|
| $\mathit{disabled}(\mathbf{0}, S)$ | | |
| $\mathit{disabled}(x^{\xi}.A, S)$ | if | $\mathit{disabled}(A, S)$ and $x \notin S$ |
| $\mathit{disabled}(a_{\mathbf{c}}.A, S)$ | if | $\mathit{disabled}(A, S)$ |
| $\mathit{disabled}(a_{\emptyset}.A, S)$ | | |
| $\mathit{disabled}(\tau_{\mathbf{c}}.A, S)$ | if | $\mathit{disabled}(A, S)$ |
| $\mathit{disabled}(\tau_{\emptyset}.A, S)$ | | |
| $\mathit{disabled}(A_1 \mid A_2, S)$ | if | $\mathit{disabled}(A_1, S)$ and $\mathit{disabled}(A_2, S)$ |
| $\mathit{disabled}(A_1 \oplus A_2, S)$ | if | $\mathit{disabled}(A_1, S)$ or $\mathit{disabled}(A_2, S)$ |
| $\mathit{disabled}(*A, S)$ | if | $\mathit{disabled}(A, S)$ |
| $\mathit{disabled}((\nu x)\, A, S)$ | if | $\mathit{disabled}(A, S\backslash\{x\})$ |
| $\mathit{disabled}(A{\uparrow_{S'}}, S)$ | if | $\mathit{disabled}(A, S\backslash S')$ |
| $\mathit{disabled}(A{\downarrow_{S'}}, S)$ | if | $\mathit{disabled}(A, S \cap S')$ |
| $\mathit{disabled}(\langle\widetilde{y}/\widetilde{x}\rangle A, S)$ | if | $\mathit{disabled}(A, \{z \mid [\widetilde{y}/\widetilde{x}]z \in S\})$ |
| $\mathit{disabled}(\mu\alpha.A, S)$ | if | $\mathit{disabled}([\mu\alpha.A/\alpha]A, S)$ |

Figure 7: The definition of $\mathit{disabled}(A, S)$

For example, a process having type $\overline{x}_{\mathbf{c}}.\,\overline{x}_{\emptyset}.\,\mathbf{0}$ implies that the process may send values on $x$ twice, and that the first send is guaranteed to succeed (i.e., the sent value will be received by some process), while there is no such guarantee for the second send.

The transition semantics of behavioral types is unchanged; The attribute $t$ is just ignored.

We revise the definitions of the subtype relation and the traces by using the following predicate $\mathit{disabled}(A, S)$. Intuitively, this means that $A$ describes a process that may get blocked without accessing any resources in $S$.

**Definition 5.5.** $\mathit{disabled}(A, S)$ is the least binary relation between extended behavioral types and sets of variables closed under the rules in Figure 7.

**Definition 5.6.** The set $\mathbf{etraces}_x(A)$ of extended traces is:

$$\{\xi_1 \cdots \xi_n \downarrow \mid \exists B. A{\downarrow_{\{x\}}} \stackrel{x^{\xi_1}}{\Longrightarrow} \cdots \stackrel{x^{\xi_n}}{\Longrightarrow} B \wedge \mathit{disabled}(B, \{x\})\}$$
$$\cup \{\xi_1 \cdots \xi_n \mid \exists B. A{\downarrow_{\{x\}}} \stackrel{x^{\xi_1}}{\Longrightarrow} \cdots \stackrel{x^{\xi_n}}{\Longrightarrow} B\}$$

Here, $A{\downarrow_{\{x\}}} \stackrel{x^{\xi_1}}{\Longrightarrow} \cdots \stackrel{x^{\xi_n}}{\Longrightarrow} B \wedge \mathit{disabled}(B, \{x\})$ means that $\xi_n$ may be the last access to $x$, so that $\downarrow$ is attached to the sequence $\xi_1 \cdots \xi_n$. By definition, $\mathbf{etraces}_x(A)$ is prefix-closed.

**Definition 5.7.** $A_1 \leq A_2$ is the largest relation on closed behavioral types that satisfies the following properties:

- If $A_1 \stackrel{l}{\longrightarrow} A_1'$ then there exists $A_2'$ such that $A_2 \stackrel{l}{\Longrightarrow} A_2'$ and $A_1' \leq A_2'$.
- $\mathit{disabled}(A_1, S)$ implies $\mathit{disabled}(A_2, S)$ for any set $S$ of variables.

Note that by the definition, $A_1 \leq A_2$ implies $\mathbf{etraces}_x(A_1) \subseteq \mathbf{etraces}_x(A_2)$.

The typing rules are the same as those in Section 3, except for the rules shown in Figure 8. The only changes are that attributes have been attached to (ET-Out) and (ET-In), and that $\mathbf{traces}_x(A{\downarrow_{\{x\}}})$ has been replaced by $\mathbf{etraces}_x(A{\downarrow_{\{x\}}})$ in (ET-NewR).

$$\frac{\Gamma \triangleright_{pl} P : A_2 \qquad \Gamma \triangleright_{pl} x : \mathbf{chan}\langle(\widetilde{y}:\widetilde{\sigma})A_1\rangle \qquad \Gamma \triangleright_{pl} \widetilde{v} : \widetilde{\sigma}}{\Gamma \triangleright_{pl} \overline{x}_t\langle\widetilde{v}\rangle.\, P : \overline{x}_t.\,(\langle\widetilde{v}/\widetilde{y}\rangle A_1 \mid A_2)} \quad \text{(ET-OUT)}$$

$$\frac{\Gamma, \widetilde{y}:\widetilde{\sigma} \triangleright_{pl} P : A_2 \qquad \Gamma \triangleright x : \mathbf{chan}\langle(\widetilde{y}:\widetilde{\sigma})A_1\rangle \qquad A_2{\downarrow_{\{\widetilde{y}\}}} \leq A_1}{\Gamma \triangleright_{pl} x_t(\widetilde{y}).\, P : x_t.\,(A_2{\uparrow_{\{\widetilde{y}\}}})} \quad \text{(ET-IN)}$$

$$\frac{\Gamma, x:\mathbf{res} \triangleright_{pl} P : A \qquad \mathbf{etraces}_x(A) \subseteq \Phi}{\Gamma \triangleright_{pl} (\mathfrak{N}^\Phi x)P : A{\uparrow_{\{x\}}}} \quad \text{(ET-NEWR)}$$

Figure 8: Typing Rules for Partial Liveness

An important invariant maintained by the typing rules is that the type of an input/output process is annotated with $\mathbf{c}$ only if the process itself is annotated with $\mathbf{c}$. For example, we cannot derive $x : \mathbf{chan}\langle\rangle \triangleright_{pl} \overline{x}_\emptyset\langle\,\rangle : \overline{x}_\mathbf{c}$.

The following theorem states the soundness of the extended type system.

**Theorem 5.8.** *If $well_annotated(P)$ and $\emptyset \triangleright_{pl} P : A$, then $P$ is partially live.*

*Proof.* We make use of three lemmas. The first two show that typing and well-annotatedness are preserved by reduction. The third means that the type of a process properly captures the possibility of the process being blocked.

- **Subject reduction.** If $\Gamma \triangleright_{pl} P : A$ and $P \stackrel{L}{\longrightarrow} Q$, then there exists some $B$ such that $\Gamma \triangleright_{pl} Q : B$ and $A \stackrel{L}{\Longrightarrow} B$. Proof: See Appendix B.
- **Well-annotatedness.** If $well_annotated(P)$ and $P \longrightarrow^* \preceq Q$, then $well_annotated(Q)$. Proof: trivial by definition of $well_annotated(P)$.
- **Disabled.** If $well_annotated(P)$ and $\Gamma \triangleright_{pl} P : A$ with $\mathbf{bool} \notin codom(\Gamma)$, then $P \not\longrightarrow$ implies $disabled(A, S)$ for any $S$. Proof: See Appendix C.

Now we are ready to prove the theorem. Suppose that $P \longrightarrow^* (\widetilde{\nu\mathfrak{N}})(\mathfrak{N}^\Phi x)Q \not\longrightarrow$ and $well_annotated(P)$, $\emptyset \triangleright_{pl} P : A$. We have to show $\downarrow\, \in \Phi$. By *subject-reduction* we obtain $\emptyset \triangleright_{pl} (\widetilde{\nu\mathfrak{N}})(\mathfrak{N}^\Phi x)Q : A'$ for some $A'$. By the inversion of the typing rules, we get $\widetilde{y}:\widetilde{\mathbf{res}}, \widetilde{z}:\widetilde{\sigma}, x:\mathbf{res} \triangleright_{pl} Q : B$ and $\mathbf{traces}_x(B) \subseteq \Phi$ for some sequence $\widetilde{\sigma}$ of channel types. (Here, $\widetilde{y}$ and $\widetilde{z}$ are the variables bound by $\mathfrak{N}$.) By *well-annotatedness* we also have $well_annotated((\widetilde{\nu\mathfrak{N}})(\mathfrak{N}^\Phi x)Q)$, which implies $well_annotated(Q)$. Thus, by *Disabled*, we get $disabled(B, S)$ for any $S$, which implies $disabled(B{\downarrow_{\{x\}}}, \{x\})$. So, we have $\downarrow\, \in \mathbf{etraces}_x(B) \subseteq \Phi$ as required. ☐

**Example 5.9.** An annotated version of Example 3.11:

$$\begin{array}{rcl}
P & = & (\nu s)\,(*s_\mathbf{c}(n,x,r).\,P_1 \mid (\mathfrak{N}^\Phi x)P_2) \\
P_1 & = & \mathbf{if}\ n=0\ \mathbf{then}\ \overline{r}_\mathbf{c}\langle\rangle \\
 & & \mathbf{else}\ (\nu r')\,(\overline{s}_\mathbf{c}\langle n-1,x,r'\rangle.\ \mid r'_\mathbf{c}().\,\mathbf{read}(x).\overline{r}_\mathbf{c}\langle\rangle) \\
P_2 & = & (\nu r)\,(\mathbf{init}(x).\overline{s}_\mathbf{c}\langle 100,x,r\rangle \mid r_\mathbf{c}().\,\mathbf{close}(x)) \\
\Phi & = & (IR^*C\downarrow)^\#
\end{array}$$

is well-annotated. Suppose

$$A_1 = \mu\alpha.(\overline{r}_{\mathbf{c}} \oplus (\nu r')\,(\langle r'/r\rangle\alpha | r'_{\mathbf{c}}.\, x^R.\overline{r}_{\mathbf{c}}) \\ \Gamma = s{:}\mathbf{chan}\langle(b{:}\mathbf{int},\, x{:}\mathbf{res},\, r{:}\mathbf{chan}\langle\rangle)\, A_1\rangle.$$

Then

$$\Gamma \triangleright P_1 : A_1 \\ \Gamma \triangleright *s_{\mathbf{c}}(n,x,r).\, P_1 : *s_{\mathbf{c}}.\,(A_1{\uparrow}_{\{n,x,r\}}) \approx *s_{\mathbf{c}} \\ \Gamma \triangleright P_2 : (\nu r)\,(x^I.A_1 | r_{\mathbf{c}}.\, x^C).$$

So long as $\mathbf{etraces}_x((\nu r)\,(x^I.A_1 \,|\, r_{\mathbf{c}}.\, x^C.)) \subseteq \Phi$, we obtain $\emptyset \triangleright P : \mathbf{0}$. □

5.2. **Type Inference.** The type inference algorithm for the extended type system is almost the same as the algorithm for the basic type system discussed in Section 4. The only changes are:

- In the constraint generaltion algorithm $PT$, attribute annotations for input and ouptut processes are propagated to types. For example, the case for output processes becomes:

$$\begin{array}{l} PT(\overline{x}^t\langle\widetilde{v}\rangle.\, P_0) = \\ \quad \mathbf{let}\;\; (\Gamma_i, \sigma_i) = PTv(v_i) \\ \quad\quad (\Gamma_0, A_0, C_0) = PT(P_0) \\ \quad\quad (\Gamma, C) = \Gamma_0 \otimes (x : \mathbf{chan}\langle(\widetilde{y} : \widetilde{\sigma})\alpha\rangle) \otimes \Gamma_1 \otimes \cdots \otimes \Gamma_n \\ \quad \mathbf{in}\; (\Gamma, \overline{x}_t.\, ([\widetilde{v}/\widetilde{y}]\alpha \,|\, A_0), C) \;\text{(where } \alpha \text{ fresh)} \end{array}$$

- The constraint $\mathbf{traces}_x(A) \subseteq \Phi$ is replaced by $\mathbf{etraces}_x(A) \subseteq \Phi$.

The second change forces us to adjust the reduction of the constraint to the reachability problem of Petri nets (recall step 3 of the algorithm in Section 4). First, we need to use $\mathbf{eptraces}(N_{A,x})$ defined below, which corresponds to $\mathbf{etraces}_x(A)$, instead of $\mathbf{ptraces}(N_{A,x})$ in the reduction.

**Definition 5.10.** $\mathbf{eptraces}(N_{A,x})$ is the set

$$\{\xi_1 \cdots \xi_k \mid m_I \stackrel{\xi_1}{\Longrightarrow} \cdots \stackrel{\xi_k}{\Longrightarrow} m'\} \cup \{\xi_1 \cdots \xi_k \downarrow \mid m_I \stackrel{\xi_1}{\Longrightarrow} \cdots \stackrel{\xi_k}{\Longrightarrow} m' \wedge \mathit{pdisabled}(m', \{x\})\}$$

where $m_I$ is the initial marking of $N_{A,x}$. $\mathit{pdisabled}(m, S)$ means that $\mathit{disabled}(A, S)$ holds for the behavioral type $A$ expressed by $m$.

Second, the construction of an automaton needs to be adjusted so that it accepts extended traces. For example, the automaton used in the explanation of Step 3-2 in Section 4 is replaced by the one that accepts $IR^*C\downarrow$.

With these changes, the validity of a constraint $\mathbf{etraces}_x(A) \subseteq \Phi$ is reduced to the reachability problem of a Petri net $N_{A,x} \,\|\, M_\Phi$ where composition of a Petri net $N_{A,x}$ and an automaton $M_\Phi$ is defined in the same manner as Definition 4.3.

**Theorem 5.11.** ***eptraces***$(N_{A,x}) \subseteq \Phi$ *if and only if no marking* $m \,|\, q$ *that satisfies the following conditions is reachable:*

- $\mathit{pdisabled}(m, \{x\})$.
- $\delta_\Phi(q, \downarrow)$ *is undefined.*

## 6. Implementation

We have implemented a prototype resource usage analyzer based on the extended type system described in Section 5. We have tested all the examples given in the present paper. The implementation can be tested at `http://www.yl.is.s.u-tokyo.ac.jp/~kohei/usage-pi/`.

The analyzer takes a pi-calculus program as an input, and uses `TyPiCal`[12] to annotate each input or output action with an attribute on whether the action is guaranteed to succeed automatically (recall the syntax of extended processes in Section 5). The annotated program is then analyzed based on the algorithm described in Section 4.

The followings are some design decisions we made in the current implementation. We restrict the resource usage specification ($\Phi$) to the regular languages, although in future we may extend it based on Remark 4.6. In Step 3-1 of the algorithm for checking $\mathbf{etraces}_x(A) \subseteq \Phi$, we blindly approximate $A$ by pushing all of its $\nu$-prefixes to the top-level. In future we might utilize an existing model checker to handle the case where $A$ is already finite. In Step 3-4 for solving the reachability problems of Petri nets, we approximate the number of tokens in each place by an element of the finite set $\{0, 1, 2, \text{“3 or more”}\}$. That approximation reduces Petri nets to finite state machines, so we can use BDD to compute an approximation of the reachable states.

Figure 9 shows a part of a successful run of the analyzer. The first process (on the second line) of the input program runs a server, which returns a new, initialized resource. We write `!` and `?` for output and input actions. The resource access specification is here expressed by the number 1 of `newR 1, x`, which refers to the built-in specification $(I(R+W)^*C\downarrow)^\#$. The second process runs infinitely many client processes, each of which sends a request for a new resource, and after receiving it, reads and closes it. The third process (on the 6th line) is a tail-recursive version of the replicated service in Example 2.9. Here, a boolean is passed as the first argument of $s$ instead of an integer, as the current system is not adapted to handle integers; it does not affect the analysis, since the system ignores the value and simply inspects both branches of the conditional. Note that the program creates infinitely many resources and has infinitely many states. The first output is the annotated version of the input program produced by `TyPiCal`, where `!!` and `??` are an output and an input with the attribute **c** (recall Section 5).

The remaining part shows the trace inclusion constraint and the constructed Petri net. The final line reports that the verification has succeeded, which implies that both the safety property (in Section 3) and the partial liveness property (in Section 5) are satisfied.

## 7. Related Work

Resource usage analysis and similar analyses have recently been studied extensively, and a variety of methods from type systems to model checking have been proposed [5, 6, 7, 11, 1, 17, 24]. However, only a few of them deal with concurrent languages. To our knowledge, none of them deal with the partial liveness property (or the total liveness property) that we discussed in Section 5. Nguyen and Rathke [20] propose an effect-type system for a kind of resource usage analysis for functional languages extended with threads and monitors. In their language, neither resources nor monitors can be created dynamically. On the other hand, our target language is $\pi$-calculus, so that our type system can be applied to programs that may create infinitely many resources (due to the existence of primitives for dynamic creation of resources: recall the example in Figure 9), and also to programs

**Input:**

```
new create,s in
  *(create?(r).newR 1,x in acc(x,init).r!(x))
| *(new r in create!(r)
     | r?(y).new c in s!(false,y,c) | s!(false,y,c)
                    | c?().c?().acc(y,close))
| *(s?(b,x,r).if b then r!()
              else acc(x,read).s!(b,x,r))
```

**Output:**

```
(*** The result of lock-freedom analysis ***)
new create, s in
  *create??(r). newR 1,x in acc(x, I). r!!(x)
| *(new r in create!!(r)
     | r??(y).new c in s!!(false,y,c) | s!!(false,y,c)
                    | c??().c??().acc(y,close))
...
(*** Constraints ***)
etrace(x,acc(x, init).(c!! & acc(x, read). $16 | $16 |
         c??. c??. acc(x, close). 0)) is included in 1
...
(*** initial marking ***)
1 * 11 | 1 * 7
(*** 14 Places ***)
0: c!!. 0
...
(*** 9 Transitions ***)
(x,close): 1*12 | 1*10 -> -1*12 | 1*13 | -1*10 | 1*1
...
No error found
```

Figure 9: A Sample Run of the Analyzer.

that use a wide range of communication and synchronization primitives. Capability-based type systems can deal with concurrency to a certain degree ([5], Section 4.2), by associating each resource with a unique capability to access the resource. The type system can control the resource access order, by ensuring the uniqueness of the capability and keeping track of what access is currently allowed by each capability. In this approach, however, resource accesses are completely serialized and programmers have to care about appropriately passing capabilities between threads. Capability-based type systems [5, 6] also require rather complex type annotations. Igarashi and Kobayashi's type system for resource usage analysis for $\lambda$-calculus [11] can be extended to deal with threads, by introducing the following typing rule:

$$\frac{\Gamma_1 \triangleright M_1 : \tau_1 \qquad \Gamma_2 \triangleright M_2 : \tau_2}{\Gamma_1 \otimes \Gamma_2 \triangleright \mathit{spawn}(M_1); M_2}$$

Here, $\Gamma_1 \otimes \Gamma_2$ describes resources that are used according to $\Gamma_1$ and $\Gamma_2$ that are used in an interleaving manner. However, it is not obvious how to accurately capture information about possible synchronizations between $M_1$ and $M_2$.

Model checking technologies [2] can of course be applicable to concurrent languages, but they suffer from the state explosion problem, especially for expressive concurrent languages like $\pi$-calculus, where resources and communication channels can be dynamically created and passed around. Appropriate abstraction must be devised for effectively performing the resource usage analysis for the $\pi$-calculus with model checking. Actually, our type-based analysis can be considered a kind of abstract model checking. The behavioral types extracted by (the first two steps of) the type inference algorithm are abstract concurrent programs, each of which captures the access behavior on each resource. Then, conformance of the abstract program with respect to the resource usage specification is checked as a model checking problem. It would be interesting to study a relationship between the abstraction through our behavioral type and the abstraction techniques for concurrent programs used in the model checking community. From that perspective, an advantage of our approach is that our type, which describes a resource-wise behavior, has much smaller state space than the whole program. In particular, if infinitely many resources are dynamically created, the whole program has infinite states, but it is often the case that our behavioral types are still finite (indeed so for the example in Figure 9). The limitation of our current analysis is that programs can be abstracted in only one way; on the other hand, the usual abstract model checking techniques refine abstraction step by step until the verification succeeds.

Technically, closest to our type system are that of Igarashi and Kobayashi [10] and that of Chaki, Rajamani, and Rehof [3]. Those type systems are developed for checking the communication behavior of a process, but by viewing a set of channels as a resource, it is possible to use those type systems directly for the resource usage analysis. We summarize below similarities and differences between those type systems [10, 3] and the type system in the present paper.

*(1) Whether types are supplied by the programmer or inferred automatically:* Types are inferred automatically in Igarashi and Kobayashi's generic type [10] and the type system of the present paper, but the type of each channel must be annotated with in Chaki et al.'s type system. The annotated type contains information about how the values (channels, in particular) sent along the channel are used by senders and receivers, and that information is used to make the type checking process compositional. For the purpose of the resource usage analysis discussed here, we think that it is a burden for programmers to declare how channels are going to be used, since their primary concern is how resources are accessed, not channels. Ideal would be to allow the user to specify some types and infer the others, like in ML. For that purpose, we need to develop an algorithm to check the conformance $A \leq B$ of an inferred type $A$ to a declared type $B$. That seems generally harder to decide than the trace inclusion constraint $\mathbf{traces}_x(A) \subseteq \Phi$, but we expect to be able to develop a sound algorithm by properly restricting the language of declared types.

*(2) The languages used as behavioral types:* All the three type systems use a fragment of CCS as the language of types to check cross-channel dependency of communications. The types in Igarashi and Kobayashi's generic type system for the $\pi$-calculus [10], however, lacks hiding, so that their type system cannot be applied to obtain precise information about resource usage. In fact, their analysis would fail even for the program in Example 2.8. Chaki et al.'s type system does use hiding, but lacks renaming as a constructor. Without the renaming constructor, the most general type does not necessarily exist, which hinders automatic type inference (recall Remark 3.12).

*(3) Algorithms for checking the conformance of inferred types with respect to specifications:* In Igarashi and Kobayashi's generic type system, how to check conformance of inferred types with respect to the user-supplied specifications was left open, and only suggested that it could be solved as a model checking problem. In Chaki et al.'s type system [3], the conformance is expressed as $A \models F$ (for checking the global behavior, where $F$ is an LTL-formula) and $A \leq A'$ (for checking the conformance of declared types with respect to inferred types). In their type checker PIPER [3], those conditions are verified using SPIN, so that $A$ is restricted to a finite-state process. Corresponding to the conformance check of the above work is the check of trace inclusion constraints $\mathbf{traces}_x(A) \subseteq \Phi$. Our algorithm based on the reduction to Petri nets works even when $A$ has infinite states.

*(4) The guaranteed properties:* Both Igarashi and Kobayashi's generic type [10] and the extended type system of the present paper can guarantee a certain lock-freedom property, that necessary communications or resource accesses are eventually performed (unless the whole process diverges), while Chaki et al.'s type system and the type system in Section 3 of the present paper do not. The guaranteed properties depend on the choice of the language of behavioral types and the subtyping relation. In the latter type systems, the ordinary simulation relation is used, so that a process's type describes only an upper-bound of the possible behavior of the process, not a lower-bound of the behavior like a certain resource access is eventually performed. Rajamani et al. [8, 23] recently introduced a more elaborate notion of simulation relation called "stuck-free conformance." Even with the stuck-free conformance relation, however, their type system [3] still cannot guarantee the lack of deadlock-freedom of a process. On the other hand, by relying on an external analysis to check deadlock-freedom, the extension in Section 5 keeps the typing rules and the subtyping relation simple, while achieving the guarantee that necessary resource accesses are eventually performed unless the whole process diverges.

Kobayashi's type systems for deadlock-freedom and livelock-freedom [16, 14, 15] and its implementation [12] form the basis of the extended type systems for partial and total liveness properties discussed in Section 5, and are used for producing well-annotated programs. Conversely, the behavioral types introduced in this paper can be used to refine the type systems for deadlock-freedom and livelock-freedom. Yoshida and Honda have also studied type systems that can guarantee certain lock-freedom properties [25, 9, 26]. So, their type systems can also be used for checking whether programs are well-annotated in the sense of Section 5.

In Section 5, we have utilized the existing analysis for deadlock-freedom to enhance the result of the resource usage analysis. Other type systems for concurrent languages may also be useful. For example, the type system for atomicity [4] can be used to infer the atomicity of a sequence of actions in a source program. By using the atomicity information, we may be able to reduce the state space of behavioral types and check the trace inclusion relation $\mathbf{etraces}_x(A) \subseteq \Phi$ more efficiently.

## 8. Conclusion

We have formalized a type system for resource usage analysis and proved its soundness. We have also developed a sound (but incomplete because of the last phase for deciding the trace inclusion relation $\mathbf{traces}_x(A) \subseteq \Phi$) algorithm for it in order to liberate programmers from the burden of writing complex type annotations. We have also implemented a prototype resource usage analyzer based on the algorithm.

There remains much future work. It is necessary to assess the effectiveness of our analysis, including the design of the type system and the algorithm for deciding the trace inclusion relation $\mathbf{traces}_x(A) \subseteq \Phi$, in more detail, and refine the analysis if necessary. It is also necessary to make the analyzer more user-friendly, by devising a method for generating comprehensive explanation of the verification result; currently, the analyzer gives only a yes/no answer. Extensions of the type system to deal with other typical synchronization primitives like join-patterns and internal choice is also left for future work.

# Appendix

## Appendix A. Properties of the Subtyping Relation

This section states and proves the properties of the subtyping relation, which are used in the proof of type soundness (Theorems 3.14 and 5.8, in particular the proofs of the lemmas in Appendices B and C), and in the type inference algorithm described in Section 4 (in particular, for transforming constraints on behavioral types).

Actually, there are two subtyping relations; the basic one in Definition 3.5 and the extended one in Definition 5.7. Since the proofs are almost the same, we state and prove the properties of the basic and extended ones simultaneously. In a few places, we have an additional condition to check for the extended case. Such places will be marked by "**Extended case only**." When we are discussing the basic case, attributes attached to actions should be ignored. We also omit them even for the extended case when they are not important.

**Lemma A.1** (Simulation relation)**.**

(1) *The subtyping relation is reflexive and transitive.*
(2) *(Simulation-up-to) Let* $\mathcal{R}$ *be a relation on behavioral types such that whenever* $A_1\mathcal{R}A_2$ *then*
   (i) $A_1 \xrightarrow{l} A_1'$ *implies* $A_2 \xRightarrow{l} A_2'$ *and* $A_1'\mathcal{R}\leq A_2'$ *for some* $A_2'$ *and*
   (ii) $\mathit{disabled}(A_1, S)$ *implies* $\mathit{disabled}(A_2, S)$*.*
   *Then* $\mathcal{R} \subseteq \leq$. *Condition (ii) is required only for the extended case.*

*Proof.* **Part 1** is trivial by the definition. To show **Part 2**, suppose $\mathcal{R}$ is a simulation up to. We show that $\mathcal{R}' = (\mathcal{R}\leq)\cup\mathcal{R}$ is a simulation, i.e., whenever $A_1\mathcal{R}'A_2$, (i) $A_1 \xrightarrow{l} A_1'$ implies $A_2 \xRightarrow{l} A_2'$ and $A_1'\mathcal{R}'A_2'$ for some $A_2'$ and (**Extended case only**) (ii) $\mathit{disabled}(A_1, S)$ implies $\mathit{disabled}(A_2, S)$. Suppose $A_1\mathcal{R}'A_2$. The case where $A_1\mathcal{R}A_2$ is trivial by the definition of the simulation-up-to. To check the other case, suppose $A_1\mathcal{R}A_3 \leq A_2$. To show (i), suppose also that $A_1 \xrightarrow{l} A_1'$. Since $\mathcal{R}$ is a simulation up to, there exists $A_3'$ such that $A_3 \xRightarrow{l} A_3'$ and $A_1'\ \mathcal{R}\leq\ A_3'$. By $A_3 \xRightarrow{l} A_3'$ and $A_3 \leq A_2$, we have $A_2'$ such that $A_2 \xRightarrow{l} A_2'$ and $A_3' \leq A_2'$. Since $\leq$ is transitive, we have $A_1'\ \mathcal{R}\leq\ A_2'$, which implies $A_1'\mathcal{R}'A_2'$.
**Extended case only:** To show (ii), suppose $\mathit{disabled}(A_1, S)$. Since $\mathcal{R}$ is a simulation up to, we have $\mathit{disabled}(A_3, S)$, which implies $\mathit{disabled}(A_2, S)$. ◻

**Lemma A.2** (Structural congruence)**.**

(1) $A|\mathbf{0} \approx A$
(2) $A|B \approx B|A$
(3) $A|(B|C) \approx (A|B)|C$
(4) $A\oplus B \approx B\oplus A$
(5) $A\oplus(B\oplus C) \approx (A\oplus B)\oplus C$
(6) $*A \approx A|*A$
(7) $(\nu x)(A|B) \approx (\nu x)A\,|\,B$ *if* $x \notin \mathbf{FV}(B)$
(8) $(\nu x)(A\oplus B) \approx (\nu x)A \oplus B$ *if* $x \notin \mathbf{FV}(B)$
(9) $[\mu\alpha.A/\alpha]A \approx \mu\alpha.A$

*Proof.* These proofs are all standard. ◻

We next show that $\leq$ is a precongruence. We first show it for some basic type constructors.

**Lemma A.3** (Precongruence, simple cases)**.** *If* $A \leq A'$ *then*

(1) $A|B \leq A'|B'$ *if* $B \leq B'$
(2) $\langle x/y\rangle A \leq \langle x/y\rangle A'$
(3) $(\nu x)A \leq (\nu x)A'$
(4) $A\uparrow_S \leq A'\uparrow_S$
(5) $A\downarrow_S \leq A'\downarrow_S$

*Proof.* These follow from the fact that the following relations are all simulations-up-to.

$$\begin{aligned}
\mathcal{R}_1 &= \{(A\,|\,B, A'\,|\,B') \mid A \leq A', B \leq B'\}\\
\mathcal{R}_2 &= \{(\langle\widetilde{y}/\widetilde{x}\rangle A, \langle\widetilde{y}/\widetilde{x}\rangle A') \mid A \leq A'\}\\
\mathcal{R}_3 &= \{((\nu x)\,A, (\nu x)\,A') \mid A \leq A'\}\\
\mathcal{R}_4 &= \{(A\uparrow_S, A'\uparrow_S) \mid A \leq A'\}\\
\mathcal{R}_5 &= \{(A\downarrow_S, A'\downarrow_S) \mid A \leq A'\}
\end{aligned}$$

□

We now show that $\leq$ is closed under arbitrary type constructors. $\mathbf{FTV}(B)$ below is the set of free (i.e., not bound by $\mu$) behavioral type variables.

**Lemma A.4** (Precongruence, general cases)**.** *If* $A \leq A'$ *and* $\mathbf{FTV}(B) \subseteq \{\alpha\}$*, then* $[A/\alpha]B \leq [A'/\alpha]B$*.*

*Proof.* Let $\mathcal{R}= \{([A/\alpha]B, [A'/\alpha]B)\}$. We will prove (i) if $[A/\alpha]B \xrightarrow{l} B_1$ then $[A'/\alpha]B \overset{l}{\Longrightarrow} B_1'$ with $B_1 \,\mathcal{R}\!\leq B_1'$, by induction on the derivation of $[A'/\alpha]B \xrightarrow{l} B_1'$. We will also prove (ii) *disabled*$([A/\alpha]B, S)$ implies *disabled*$([A'/\alpha]B, S)$, by induction on the structure of $B$ in the extended case. In other words, $\mathcal{R}$ is a simulation-up-to. Hence (Lemma A.1.2) it is in $\leq$.

We start with (i), with case analysis on the last rule used. If $B = \alpha$, then the required condition follows immediately from $A \leq A'$. So we consider the case $B \neq \alpha$ below.

(1) Case (Tr-Act). In this case, $B = l.B_x$, so
$$[A/\alpha]B = l.[A/\alpha]B_x \xrightarrow{l} [A/\alpha]B_x = B_1.$$
We also have
$$[A'/\alpha]B = l.[A/\alpha]B_x \xrightarrow{l} [A'/\alpha]B_x = B_1'.$$
By construction of $\mathcal{R}$, we have $B_1 \,\mathcal{R}\, B_1' \leq B_1'$ as required.

(2) Case (Tr-Par1). We show only the left case. $B = B_x|B_y$ and we assumed $[A/\alpha]B_x \xrightarrow{l} B_{x1}$ to make
$$[A/\alpha]B = [A/\alpha]B_x|[A/\alpha]B_y \xrightarrow{l} B_{x1}|[A/\alpha]B_y = B_1.$$
By the induction hypothesis, $[A'/\alpha]B_x \overset{l}{\Longrightarrow} B_{x1}'$ with $B_{x1} \,\mathcal{R}\!\leq B_{x1}'$. (Note that $\alpha$ is not free in $B_{x1}$ or $B_{x1}'$. ) That gives
$$[A'/\alpha]B = [A'/\alpha]B_x|[A'/\alpha]B_y \overset{l}{\Longrightarrow} B_{x1}'|[A'/\alpha]B_y = B_1'.$$
It remains to prove $B_1 \,\mathcal{R}\!\leq B_1'$. By the condition $B_{x1} \,\mathcal{R}\!\leq B_{x1}'$, there exists $C$ such that
$$B_{x1} = [A/\alpha]C \qquad [A'/\alpha]C \leq B_{x1}'$$

So, we get:

$$\begin{aligned} B_1 &= [A/\alpha](C \mid B_y)\ \mathcal{R}\ [A'/\alpha](C \mid B_y) \\ &= [A'/\alpha]C \mid [A'/\alpha]B_y \ \leq\ B'_{x1} \mid [A'/\alpha]B_y = B'_1. \end{aligned}$$

Here, we have used Lemma A.3, Part 1.

(3) Case (Tr-Par2). We show only the left case. $B = B_x|B_y$ and we assumed $[A/\alpha]B_x \xrightarrow{x} B_{x1}$ and $[A/\alpha]B_y \xrightarrow{\overline{y}} B_{y1}$ to make

$$[A/\alpha]B = [A/\alpha]B_x|[A/\alpha]B_y \xrightarrow{\{x,\overline{y}\}} B_{x1}|B_{y1} = B_1.$$

By the induction hypothesis, $[A'/\alpha]B_x \stackrel{x}{\Longrightarrow} B'_{x1}$ and $[A'/\alpha]B_y \stackrel{x}{\Longrightarrow} B'_{y1}$ with $B_{x1}\ \mathcal{R}{\leq}\ B'_{x1}$ and $B_{y1}\ \mathcal{R}{\leq}\ B'_{y1}$. That gives

$$[A'/\alpha]B = [A'/\alpha]B_x \mid [A'/\alpha]B_y \stackrel{\{x,\overline{y}\}}{\Longrightarrow} B'_{x1} \mid B'_{y1} = B'_1.$$

It remains to prove $B_1\ \mathcal{R}{\leq}\ B'_1$. From $B_{x1}\ \mathcal{R}{\leq}\ B'_{x1}$ and $B_{y1}\ \mathcal{R}{\leq}\ B'_{y1}$, there exist $C_x$ and $C_y$ such that

$$\begin{array}{ll} B_{x1} = [A/\alpha]C_x & [A'/\alpha]C_x \ \leq\ B'_{x1} \\ B_{y1} = [A/\alpha]C_y & [A'/\alpha]C_y \ \leq\ B'_{y1} \end{array}$$

Hence, $B_1 = [A/\alpha](C_x \mid C_y)\ \mathcal{R}\ [A'/\alpha](C_x \mid C_y)\ \leq\ B'_1$.

(4) Cases (Tr-Com) and (Tr-Or). These cases follow immediately from the induction hypothesis.

(5) Case (Tr-Rep). Then $B = *B_x$ and $[A/\alpha]B = *[A/\alpha]B_x \xrightarrow{l}$. $[A/\alpha]B \xrightarrow{l} B_1$ must have been derived from

$$[A/\alpha](B_x \mid *B_x) = [A/\alpha]B_x \mid *[A/\alpha]B_x \xrightarrow{l} B_1.$$

By the induction hypothesis, there exists $B'_1$ such that

$$B_1\ \mathcal{R}{\leq}\ B'_1 \quad \text{and} \quad [A'/\alpha](B_x \mid *B_x) \stackrel{l}{\Longrightarrow} B'_1\ .$$

Using (Tr-Rep), we get $[A'/\alpha]B \stackrel{l}{\Longrightarrow} B'_1$ as required.

(6) Case (Tr-Rec). Then, we have $B = \mu\beta.B_x$ to make

$$[A/\alpha]B = \mu\beta.[A/\alpha]B_x \xrightarrow{l} B_1$$

where we assumed $[\mu\beta.[A/\alpha]B_x/\beta][A/\alpha]B_x \xrightarrow{l} B_1$. But $\beta$ does not clash with $A$ or $\alpha$ so these two substitutions swap around, giving

$$[A/\alpha][\mu\beta.B_x/\beta]B_x \xrightarrow{l} B_1.$$

By the induction hypothesis,

$$[A'/\alpha][\mu\beta.B_x/\beta]B_x \stackrel{l}{\Longrightarrow} B'_1$$

with $B_1\ \mathcal{R}{\leq}\ B'_1$. Hence

$$[A'/\alpha]B = \mu\beta.[A'/\alpha]B_x \stackrel{l}{\Longrightarrow} B'_1$$

as required.

(7) Case (Tr-Rename). Then, $B = \langle\widetilde{y}/\widetilde{x}\rangle B_x$. $[A/\alpha]B \stackrel{[\widetilde{y}/\widetilde{x}]l}{\longrightarrow} \langle\widetilde{y}/\widetilde{x}\rangle B_{x1} = B_1$ must have been derived from $[A/\alpha]B_x \stackrel{l}{\longrightarrow} B_{x1}$. From the induction hypothesis, we get

$$[A'/\alpha]B_x \stackrel{l}{\longrightarrow} B'_{x1} \qquad B_{x1}\,\mathcal{R}\!\leq\, B'_{x1}.$$

Let $B'_1 = \langle\widetilde{y}/\widetilde{x}\rangle B'_{x1}$. It remains to prove $B_1\,\mathcal{R}\!\leq\, B'_1$. By $B_{x1}\,\mathcal{R}\!\leq\, B'_{x1}$, there exists $C$ such that

$$B_{x1} = [A/\alpha]C \qquad [A'/\alpha]C \,\leq\, B'_{x1}.$$

So, we have:

$$\begin{aligned} B_1 &= [A/\alpha]\langle\widetilde{y}/\widetilde{x}\rangle C \;\mathcal{R}\; [A'/\alpha]\langle\widetilde{y}/\widetilde{x}\rangle C \\ &= \langle\widetilde{y}/\widetilde{x}\rangle[A'/\alpha]C \,\leq\, \langle\widetilde{y}/\widetilde{x}\rangle B'_{x1} = B'_1. \end{aligned}$$

Here, we used the fact that $\leq$ is preserved by $\langle\widetilde{y}/\widetilde{x}\rangle$ (Lemma A.3, Part 2).

(8) Cases (Tr-Hiding), (Tr-Exclude), and (Tr-Project): Similar to (Tr-Rename). We use the fact that $\leq$ is preserved by $\nu$, $\cdot{\downarrow_S}$, and $\cdot{\uparrow_S}$ (Lemma A.3).

**Extended case only:** In addition we need to show that *disabled*$([A/\alpha]B, S)$ implies *disabled*$([A'/\alpha]B, S)$. This follows by straightforward induction on the structure of $B$. □

**Lemma A.5** (Substitution)**.**

(1) $\langle\widetilde{y}/\widetilde{x}\rangle \boldsymbol{0} \approx \boldsymbol{0}$
(2) $\langle\widetilde{y}/\widetilde{x}\rangle(a.A) \approx ([\widetilde{y}/\widetilde{x}]a).\langle\widetilde{y}/\widetilde{x}\rangle A$
(3) $\langle\widetilde{y}/\widetilde{x}\rangle(z^{\xi}.A) \approx ([\widetilde{y}/\widetilde{x}]z)^{\xi}.\langle\widetilde{y}/\widetilde{x}\rangle A$
(4) $\langle\widetilde{y}/\widetilde{x}\rangle(A|B) \approx \langle\widetilde{y}/\widetilde{x}\rangle A \mid \langle\widetilde{y}/\widetilde{x}\rangle B$
(5) $\langle\widetilde{y}/\widetilde{x}\rangle(A\oplus B) \approx \langle\widetilde{y}/\widetilde{x}\rangle A \;\oplus\; \langle\widetilde{y}/\widetilde{x}\rangle B$
(6) $\langle\widetilde{y}/\widetilde{x}\rangle(*A) \approx *(\langle\widetilde{y}/\widetilde{x}\rangle A)$
(7) $\langle\widetilde{y}/\widetilde{x}\rangle\langle b/a\rangle A \approx \langle[\widetilde{y}/\widetilde{x}]b/a\rangle\langle\widetilde{y}/\widetilde{x}\rangle A$ *if* $\mathit{target}(a)\cap\{\widetilde{x},\widetilde{y}\}=\emptyset$
(8) $\langle\widetilde{y}/\widetilde{x}\rangle(\nu z)A \approx (\nu z)(\langle\widetilde{y}/\widetilde{x}\rangle A)$ *if* $\{z\}\cap\{x,y\}=\emptyset$
(9) $\langle\widetilde{y}/\widetilde{x}\rangle(A{\uparrow_S}) \approx (\langle\widetilde{y}/\widetilde{x}\rangle A){\uparrow_S}$, *and*
$\langle\widetilde{y}/\widetilde{x}\rangle(A{\downarrow_S}) \approx \langle\widetilde{y}/\widetilde{x}\rangle A{\downarrow_S} \approx A{\downarrow_S}$,
*if* $S\cap\{x,y\}=\emptyset$
(10) $\langle\widetilde{y}/\widetilde{x}\rangle(A{\uparrow_S}) \approx A{\uparrow_S}$, *if* $\{\widetilde{x}\} \subseteq S$

*Proof.* Most parts are straightforward, although **Part 4** is non-obvious in the case of labels $\{x,\overline{y}\}$. For Part 4, we construct a relation $\mathcal{S} = \{(\langle\widetilde{y}/\widetilde{x}\rangle(A|B), \langle\widetilde{y}/\widetilde{x}\rangle A|\langle\widetilde{y}/\widetilde{x}\rangle B)\}$ and prove $\mathcal{S}$ and $\mathcal{S}^{-1}$ are simulations. The interesting case is when we infer

$$\langle\widetilde{y}/\widetilde{x}\rangle A|\langle\widetilde{y}/\widetilde{x}\rangle B \stackrel{\tau}{\to} \langle\widetilde{y}/\widetilde{x}\rangle A'|\langle\widetilde{y}/\widetilde{x}\rangle B'$$

from

$$A \stackrel{z_1}{\to} A' \qquad B \stackrel{\overline{z_2}}{\to} B' \qquad [\widetilde{y}/\widetilde{x}]z_1 = [\widetilde{y}/\widetilde{x}]z_2.$$

This gives

$$A|B \stackrel{\{z_1,\overline{z_2}\}}{\to} A'|B'.$$

Hence

$$\langle\widetilde{y}/\widetilde{x}\rangle(A|B) \stackrel{\{[\widetilde{y}/\widetilde{x}]z_1,[\widetilde{y}/\widetilde{x}]\overline{z_2}\}}{\longrightarrow} \langle\widetilde{y}/\widetilde{x}\rangle(A'|B').$$

And hence as required

$$\langle\widetilde{y}/\widetilde{x}\rangle(A|B) \stackrel{\tau}{\to} \langle\widetilde{y}/\widetilde{x}\rangle(A'|B').$$

**Part 9.** Here we construct $\mathcal{S} = \{(\,\langle\widetilde{y}/\widetilde{x}\rangle(A{\uparrow_S}),\ (\langle\widetilde{y}/\widetilde{x}\rangle A){\uparrow_S}\,)\}$ where $S$ does not clash with $\{\widetilde{x},\widetilde{y}\}$, and we prove that $\mathcal{S}$ and $\mathcal{S}^{-1}$ are simulations. We focus on two cases.

(1) Suppose $(\langle\widetilde{y}/\widetilde{x}\rangle A)\!\uparrow_S \xrightarrow{[\widetilde{y}/\widetilde{x}]l} (\langle\widetilde{y}/\widetilde{x}\rangle A')\!\uparrow_S$ is inferred from

$$A \xrightarrow{l} A' \quad \text{and} \quad \mathit{target}([\widetilde{y}/\widetilde{x}]l)\cap S = \emptyset.$$

We must infer that $\langle\widetilde{y}/\widetilde{x}\rangle(A\!\uparrow_S) \xrightarrow{[\widetilde{y}/\widetilde{x}]l} \langle\widetilde{y}/\widetilde{x}\rangle(A'\!\uparrow_S)$. This requires $\mathit{target}(l)\cap S = \emptyset$, which we prove as follows. It is assumed that $S$ does not clash, so $\{\widetilde{x},\widetilde{y}\}\cap S = \emptyset$. We also have $\mathit{target}([\widetilde{y}/\widetilde{x}]l)\cap S = \emptyset$, and so $[\widetilde{y}/\widetilde{x}](\mathit{target}(l))\cap S = \emptyset$. Let $T = \mathit{target}(l)$. Suppose $z \in T$. Then either $z \in \widetilde{x}$ so $z \notin S$, or $z \in \widetilde{y}$ so $z \notin S$, or $z \notin \{\widetilde{x},\widetilde{y}\}$ so $z \in [\widetilde{y}/\widetilde{x}]T$ so $z \notin S$. In all cases $z \notin S$, so $T\cap S = \emptyset$ as required.

(2) Suppose $(\langle\widetilde{y}/\widetilde{x}\rangle A)\!\uparrow_S \xrightarrow{\tau} (\langle\widetilde{y}/\widetilde{x}\rangle A')\!\uparrow_S$ is inferred from $A \xrightarrow{l} A'$ and $\mathit{target}([\widetilde{y}/\widetilde{x}]l) \subseteq S$. We must infer $\langle\widetilde{y}/\widetilde{x}\rangle(A\!\uparrow_S) \xrightarrow{\tau} \langle\widetilde{y}/\widetilde{x}\rangle(A'\!\uparrow_S)$. This requires $\mathit{target}(l) \subseteq S$, which we prove as follows. Once again let $T = \mathit{target}(l)$. We have $\{\widetilde{x},\widetilde{y}\}\cap S = \emptyset$ and $[\widetilde{y}/\widetilde{x}]T \subseteq S$. Suppose $z \in T$. Then $[\widetilde{y}/\widetilde{x}]z \in [\widetilde{y}/\widetilde{x}]T$, and $[\widetilde{y}/\widetilde{x}]z \in S$. Either $z \in \widetilde{x}$ so $y \in S$, which is a contradiction. Or $z \notin \widetilde{x}$, so $[\widetilde{y}/\widetilde{x}]z = z \in S$. Hence $T \subseteq S$ as required. □

**Lemma A.6** (Exclusion and Projection)**.**

| | | | |
|---|---|---|---|
| (1) | $\mathbf{0}\!\uparrow_S\approx\mathbf{0}$ | $\mathbf{0}\!\downarrow_S\approx\mathbf{0}$ | |
| (2) | $(a_t.A)\!\uparrow_S\approx a_t.(A\!\uparrow_S)$ | $(a_t.A)\!\downarrow_S\approx\tau_t.(A\!\downarrow_S)$ | *if* $\mathit{target}(a)\cap S=\emptyset$ |
| (3) | $(a_t.A)\!\uparrow_S\approx\tau_t.A\!\uparrow_S$ | $(a_t.A)\!\downarrow_S\approx a_t.A\!\downarrow_S$ | *if* $\mathit{target}(a)\subseteq S$ |
| (4) | $(z^\xi.A)\!\uparrow_S\approx z^\xi.(A\!\uparrow_S)$ | $(z^\xi.A)\!\downarrow_S\approx\tau_c.A\!\downarrow_S$ | *if* $\mathit{target}(z^\xi)\cap S=\emptyset$ |
| (5) | $(z^\xi.A)\!\uparrow_S\approx\tau_c.A\!\uparrow_S$ | $(z^\xi.A)\!\downarrow_S\approx z^\xi.(A\!\downarrow_S)$ | *if* $\mathit{target}(z^\xi)\subseteq S$ |
| (6) | $(A\mid B)\!\uparrow_S\approx A\!\uparrow_S \mid B\!\uparrow_S$ | $(A\mid B)\!\downarrow_S\approx A\!\downarrow_S \mid B\!\downarrow_S$ | |
| (7) | $(A\oplus B)\!\uparrow_S\approx A\!\uparrow_S\oplus B\!\uparrow_S$ | $(A\oplus B)\!\downarrow_S\approx A\!\downarrow_S\oplus B\!\downarrow_S$ | |
| (8) | $(*A)\!\uparrow_S\approx *(A\!\uparrow_S)$ | $(*A)\!\downarrow_S\approx *(A\!\downarrow_S)$ | |
| (9) | $(A\!\uparrow_S)\!\uparrow_T\approx A\!\uparrow_{S\cup T}$ | $(A\!\downarrow_S)\!\downarrow_T\approx A\!\downarrow_{S\cap T}$ | |
| (10) | $A\!\uparrow_S\approx A$ | $A\!\downarrow_S\leq\mathbf{0}$ | *if* $\mathbf{FV}(A)\cap S=\emptyset$ |
| (11) | $A\!\uparrow_S\leq\mathbf{0}$ | $A\!\downarrow_S\approx A$ | *if* $\mathbf{FV}(A)\subseteq S$ |

*Proof.* Straightforward. □

**Lemma A.7** (Simulation)**.**

(1) *If* $A_1 \leq A_2$ *then* $\boldsymbol{traces}_x(A_1) \subseteq \boldsymbol{traces}_x(A_2)$ *for any* $x$.
(2) *If* $A \xrightarrow{\{x,\overline{y}\}} A'$ *then* $A \xrightarrow{x}\xrightarrow{\overline{y}} A'$.
(3) $A \leq A\oplus B$
(4) $A\oplus A \leq A$
(5) $A \leq A\!\uparrow_S \mid A\!\downarrow_S$
(6) *If* $[B/\alpha]A \leq B$ *then* $\mu\alpha.A \leq B$
(7) $B_1\oplus B_2 \leq A$ *if and only if* $B_1 \leq A$ *and* $B_2 \leq A$

*Proof.* These proofs are largely standard.

**Part 1** follows immediately from the definitions of subtyping and traces.

**Part 6.** Suppose $[B/\alpha]A \leq B$. Let $\mathcal{R}$ be

$$\{([\mu\alpha.A/\alpha]A', [B/\alpha]A') \mid \mathbf{FTV}(A') = \{\alpha\}\}.$$

By Lemma A.3.2, It suffices to prove that $\mathcal{R}$ is a simulation up to.

Suppose that $[\mu\alpha.A/\alpha]A' \;\mathcal{R}\; [B/\alpha]A'$ and $[\mu\alpha.A/\alpha]A' \xrightarrow{l} A''$. We show that there exists $B'$ such that $[B/\alpha]A' \xRightarrow{l} B'$ and $A'' \;\mathcal{R}\!\leq\; B'$ by induction on the derivation of $[\mu\alpha.A/\alpha]A' \xrightarrow{l} A''$, with case analysis on the last rule used. We show main cases; the other cases are similar or straightforward.

- Case (TR-Act): $[\mu\alpha.A/\alpha]A' \xrightarrow{l} A''$ is derived from $l.\,[\mu\alpha.A/\alpha]A_1 \xrightarrow{l} [\mu\alpha.A/\alpha]A_1$ where $A' = l.\,A_1$ and $A'' = [\mu\alpha.A/\alpha]A_1$. Thus, $[B/\alpha]A' = l.\,[B/\alpha]A_1 \xrightarrow{l} [B/\alpha]A_1$.
- Case (TR-Par1): $[\mu\alpha.A/\alpha]A' \xrightarrow{l} A''$ is derived from $[\mu\alpha.A/\alpha]A_1 \xrightarrow{l} A'_1$ where $A' = A_1 \,|\, A_2$ and $A'' = A'_1 \,|\, [\mu\alpha.A/\alpha]A_2$. By the induction hypothesis, there exists $B'_1$ such that $[B/\alpha]A_1 \stackrel{l}{\Longrightarrow} B'_1$ and $A'_1 \;\mathcal{R}{\leq}\; B'_1$. Thus, we have $[B/\alpha]A' \stackrel{l}{\Longrightarrow} B'_1 \,|\, [B/\alpha]A_2$. It remains to show $A'' = A'_1 \,|\, [\mu\alpha.A/\alpha]A_2 \;\mathcal{R}{\leq}\; B'_1 \,|\, [B/\alpha]A_2$. From $A'_1 \;\mathcal{R}{\leq}\; B'_1$, we get
$$A'_1 = [\mu\alpha.A/\alpha]C \qquad [B/\alpha]C \,\leq\, B'_1$$
for some $C$. So,
$$\begin{array}{l} A'' = A'_1 \,|\, [\mu\alpha.A/\alpha]A_2 = [\mu\alpha.A/\alpha](C \,|\, A_2) \\ \mathcal{R}\; [B/\alpha](C \,|\, A_2) = [B/\alpha]C \,|\, [B/\alpha]A_2 \,\leq\, B'_1 \,|\, [B/\alpha]A_2 \end{array}$$
- Case (TR-Par2): $[\mu\alpha.A/\alpha]A' \xrightarrow{\{x,\overline{y}\}} A''$ is derived from $[\mu\alpha.A/\alpha]A_1 \xrightarrow{x} A'_1$ and $[\mu\alpha.A/\alpha]A_2 \xrightarrow{\overline{y}} A'_2$ where $A' = A_1 \,|\, A_2$ and $A'' = A'_1 \,|\, A'_2$. From the induction hypothesis, there exist $B'_1$ and $B'_2$ such that $[B/\alpha]A_1 \stackrel{x}{\Longrightarrow} B'_1$ and $A'_1 \;\mathcal{R}{\leq}\; B'_1$ and $[B/\alpha]A_2 \stackrel{\overline{y}}{\Longrightarrow} B'_2$ and $A'_2 \;\mathcal{R}{\leq}\; B'_2$. Thus, we have $[B/\alpha]A' \stackrel{\{x,\overline{y}\}}{\Longrightarrow} B'_1 \,|\, B'_2$. From $A'_1 \;\mathcal{R}{\leq}\; B'_1$ and $A'_2 \;\mathcal{R}{\leq}\; B'_2$, we get $A'_1 \,|\, A'_2 \;\mathcal{R}{\leq}\; B'_1 \,|\, B'_2$ as required.
- Case (TR-Rec):
  - Case $A' = \mu\beta.A_1$: $[\mu\alpha.A/\alpha]A' \xrightarrow{l} A''$ is derived from
  $$\begin{array}{l} [\mu\alpha.A/\alpha][\mu\beta.A_1/\beta]A_1 \\ = [\mu\beta.[\mu\alpha.A/\alpha]A_1/\beta][\mu\alpha.A/\alpha]A_1 \xrightarrow{l} A''. \end{array}$$
  Here, we assumed without loss of generality that $\beta$ is not free in $A$ and $B$. Thus, by the induction hypothesis, there exists $B'$ such that
  $$[\mu\beta.[B/\alpha]A_1/\beta][B/\alpha]A_1 = [B/\alpha][\mu\beta.A_1/\beta]A_1 \stackrel{l}{\Longrightarrow} B'$$
  and $A'' \;\mathcal{R}{\leq}\; B'$. Using (Tr-Rec), we obtain $[B/\alpha]A' = \mu\beta.[B/\alpha]A_1 \stackrel{l}{\Longrightarrow} B'$ as required.
  - Case $A' = \alpha$: $[\mu\alpha.A/\alpha]A'$ is equal to $\mu\alpha.A$. From $\mu\alpha.A \,\leq\, B$, there exists $B'$ such that $B \stackrel{l}{\Longrightarrow} B'$ and $A'' \,\leq\, B'$ as required.

**Extended case only:** We also need to prove that
$$\mathit{disabled}([\mu\alpha.A/\alpha]A', S) \quad \text{implies} \quad \mathit{disabled}([B/\alpha]A', S) \quad \text{for any} \quad A'\ .$$
This is proved by induction on the derivation of $\mathit{disabled}([\mu\alpha.A/\alpha]A', S)$. We show the only non-trivial case, where $\mathit{disabled}([\mu\alpha.A/\alpha]A', S)$ has been derived by using the last rule in Figure 7. The other cases follow immediately from the induction hypothesis.

There are two cases to consider.

- Case where $A' = \alpha$: Then, $[\mu\alpha.A/\alpha]A' = \mu\alpha.A$ and *disabled*$(\mu\alpha.A, S)$ must have been deduced from
*disabled*$([\mu\alpha.A/\alpha]A, S)$. By the induction hypothesis, we have *disabled*$([B/\alpha]A, S)$. By the assumption $[B/\alpha]A \leq B$, we have *disabled*$(B, S)$ as required (note that $[B/\alpha]A' = B$ in this case).
- Case where $A' = \mu\beta.C$. Let $C'$ be $[\mu\alpha.A/\alpha]C$. Then, $[\mu\alpha.A/\alpha]A' = \mu\beta.C'$, and *disabled*$(\mu\beta.C', S)$ must have been derived from *disabled*$([\mu\beta.C'/\beta]C', S)$. Here, we note
$$[\mu\beta.C'/\beta]C' = [\mu\alpha.A/\alpha][\mu\beta.C/\beta]C.$$
So, from the induction hypothesis, we get
*disabled*$([B/\alpha][\mu\beta.C/\beta]C, S)$, i.e.,
$$\mathit{disabled}([\mu\beta.[B/\alpha]C/\beta][B/\alpha]C, S).$$
By using the last rule of Figure 7, we get *disabled*$([B/\alpha]A', S)$ as required. □

## Appendix B. Proof of the Subject Reduction Property

In this section, we prove the subject reduction property used in the proofs of Theorems 3.14 and 5.8. As in Appendix A, we prove it for the basic and extended cases simultaneously.

**Lemma B.1** (Weakening)**.** (1) *If* $\Gamma \triangleright v : \sigma$ *and* $x \notin \mathit{dom}(\Gamma)$*, then* $\Gamma, x : \sigma' \triangleright v : \tau$*.*
(2) *If* $\Gamma \triangleright P : A$ *and* $x \notin \mathbf{FV}(P)$ *and* $x$ *not in* $\mathit{dom}(\Gamma)$ *or* $\mathbf{FV}(A)$ *then* $\Gamma,\, x : \sigma \triangleright P : A$*.*

*Proof.* Part 1 is straightforward. Part 2 is proved by straightforward induction on the derivation of $\Gamma \triangleright P : A$. □

**Lemma B.2** (Judgement substitution)**.**
(1) *(For values) If* $\Gamma, \widetilde{x} : \widetilde{\sigma} \triangleright y : \sigma$ *and* $\Gamma \triangleright \widetilde{v} : \widetilde{\sigma}$ *then* $\Gamma \triangleright [\widetilde{v}/\widetilde{x}]y : \sigma$*.*
(2) *(For processes) If* $\Gamma, \widetilde{x} : \widetilde{\sigma} \triangleright P : A$ *and* $\Gamma \triangleright \widetilde{v} : \widetilde{\sigma}$ *then* $\Gamma \triangleright [\widetilde{v}/\widetilde{x}]P : \langle\widetilde{v}/\widetilde{x}\rangle A$*.*

*Proof.* **Part 1.** Either $y = x_i$ for some $i$, in which case $[\widetilde{v}/\widetilde{x}]y = v_i$ and $\sigma = \sigma_i$, so that the result follows from $\Gamma \triangleright \widetilde{v} : \widetilde{\sigma}$. Or $y \notin \widetilde{x}$, in which case $[\widetilde{v}/\widetilde{x}]y = y$ and $y : \sigma$ is in $\Gamma$. We remark that types $\sigma$ never have free names.

**Part 2.** By induction on the derivation of $\Gamma \triangleright P : A$. Most cases follow straightforwardly on Lemma A.5. We consider four particular cases.

(1) Case (T-Sub), where $\Gamma, \widetilde{x} : \widetilde{\tau} \triangleright P : A$ is inferred from
$$\Gamma, \widetilde{x} : \widetilde{\tau} \triangleright P : A' \qquad A' \leq A$$
From the induction hypothesis, $\Gamma \triangleright [\widetilde{v}/\widetilde{x}]P : \langle\widetilde{v}/\widetilde{x}\rangle A'$. By Lemma A.3.2 and assumption $A' \leq A$ we get $\langle\widetilde{v}/\widetilde{x}\rangle A' \leq \langle\widetilde{v}/\widetilde{x}\rangle A$, and hence as required $\Gamma \triangleright [\widetilde{v}/\widetilde{x}]P : \langle\widetilde{v}/\widetilde{x}\rangle A$.
(2) Case (T-NewR), where $\Gamma, \widetilde{x} : \widetilde{\tau} \triangleright (\mathfrak{N}^{\Phi} z)P : A{\uparrow_{\{z\}}}$ is inferred from
$$\Gamma, \widetilde{x} : \widetilde{\tau}, z : \mathbf{res} \triangleright P : A \qquad \mathbf{traces}_z(A) \subseteq \Phi$$
Assume by alpha-renaming that $z$ does not clash with $\widetilde{x}$ or $\widetilde{v}$. From Lemma A.5.9 we get $A{\downarrow_{\{z\}}} \approx (\langle\widetilde{v}/\widetilde{x}\rangle A){\downarrow_{\{z\}}}$, giving $\mathbf{traces}_z(A) = \mathbf{traces}_z(\langle\widetilde{v}/\widetilde{x}\rangle A)$ and hence $\mathbf{traces}_z(\langle\widetilde{v}/\widetilde{x}\rangle A) \subseteq \Phi$. From $\Gamma \triangleright \widetilde{v} : \widetilde{\tau}$ and Lemma B.1, we get $\Gamma, z : \mathbf{res} \triangleright \widetilde{v} : \widetilde{\tau}$. So, by the induction hypothesis, $\Gamma, z : \mathbf{res} \triangleright [\widetilde{v}/\widetilde{x}]P : \langle\widetilde{v}/\widetilde{x}\rangle A$. These two together give
$$\Gamma \triangleright (\mathfrak{N}^{\Phi} z)[\widetilde{v}/\widetilde{x}]P : (\langle\widetilde{v}/\widetilde{x}\rangle A){\uparrow_{\{z\}}}.$$

For the process $(\mathfrak{N}^{\Phi} z)[\widetilde{v}/\widetilde{x}]P$, we can push the substitution out by definition of the substitution operator and because $z \notin \{\widetilde{x}, \widetilde{v}\}$. For the behavior $(\langle\widetilde{v}/\widetilde{x}\rangle A)\uparrow_{\{z\}}$ we use Lemma A.5.9 to push it out. Hence as required,

$$\Gamma \triangleright [\widetilde{v}/\widetilde{x}](\mathfrak{N}^{\Phi} z)P : \langle\widetilde{v}/\widetilde{x}\rangle(A\uparrow_{\{z\}}).$$

**Extended case only:** Just replace **traces** with **etraces** in the above reasoning.

(3) Case (T-Out), where $\Gamma, \widetilde{x} : \widetilde{\tau} \triangleright \overline{z}\langle w\rangle.\, P : \overline{z}.\, (\langle\widetilde{w}/\widetilde{y}\rangle A_1 | A_2)$ is inferred from

$$\begin{array}{c}\Gamma, \widetilde{x} : \widetilde{\tau} \triangleright P : A_2 \qquad \Gamma, \widetilde{x} : \widetilde{\tau} \triangleright \widetilde{w} : \widetilde{\sigma} \\ \Gamma, \widetilde{x} : \widetilde{\tau} \triangleright z : \mathbf{chan}\langle(\widetilde{y} : \widetilde{\sigma})A_1\rangle\end{array}$$

Part 1 implies $\Gamma \triangleright [\widetilde{v}/\widetilde{x}]\widetilde{w} : \widetilde{\sigma}$ and $\Gamma \triangleright [\widetilde{v}/\widetilde{x}]z : \mathbf{chan}\langle(\widetilde{y} : \widetilde{\sigma})A_1\rangle$. From the induction hypothesis, we get $\Gamma \triangleright [\widetilde{v}/\widetilde{x}]P : \langle\widetilde{v}/\widetilde{x}\rangle A_2$. These three give

$$\Gamma \triangleright \overline{[\widetilde{v}/\widetilde{x}]z}\langle[\widetilde{v}/\widetilde{x}]\widetilde{w}\rangle.\, [\widetilde{v}/\widetilde{x}]P : \overline{[\widetilde{v}/\widetilde{x}]z}.\, (\langle[\widetilde{v}/\widetilde{x}]\widetilde{w}/\widetilde{y}\rangle A_1 | \langle\widetilde{v}/\widetilde{x}\rangle A_2)$$

For the process we push the substitution out by definition of the substitution operator. For the behavior we push it out using several parts of Lemma A.5.

(4) Case (T-In), where $\Gamma, \widetilde{x} : \widetilde{\tau} \triangleright z(\widetilde{y}).\, P : z.\, (A_2\uparrow_{\{\widetilde{y}\}})$ is inferred from

$$\begin{array}{c}\Gamma, \widetilde{y} : \widetilde{\sigma}, \widetilde{x} : \widetilde{\tau} \triangleright P : A_2 \qquad \Gamma, \widetilde{x} : \widetilde{\tau} \triangleright z : \mathbf{chan}\langle(\widetilde{y} : \widetilde{\sigma})A_1\rangle \\ A_2\downarrow_{\{\widetilde{y}\}} \le A_1\end{array}$$

We use three deductions. First from Part 1 we get $\Gamma \triangleright [\widetilde{v}/\widetilde{x}]z : \mathbf{chan}\langle(\widetilde{y} : \widetilde{\sigma})A_1\rangle$. Second, from assumption $A_2\downarrow_{\{\widetilde{y}\}} \le A_1$ and Lemma A.3.2 we get $\langle\widetilde{v}/\widetilde{x}\rangle(A_2\downarrow_{\{\widetilde{y}\}}) \le \langle\widetilde{v}/\widetilde{x}\rangle A_1$. The substitution on the right disappears because $\mathbf{FV}(A_1) \subseteq \{\widetilde{y}\}$ and we can assume by alpha-renaming that $\widetilde{y}$ does not clash with $\{\widetilde{x}, \widetilde{v}\}$. The substitution on the left can be pushed inside by Lemma A.5.9. These together give $(\langle\widetilde{v}/\widetilde{x}\rangle A_2)\downarrow_{\{\widetilde{y}\}} \le A_1$. And third, from the induction hypothesis we get $\Gamma, \widetilde{y} : \widetilde{\sigma} \triangleright [\widetilde{v}/\widetilde{x}]P : \langle\widetilde{v}/\widetilde{x}\rangle A_2$. These three give

$$\Gamma \triangleright [\widetilde{v}/\widetilde{x}]z(\widetilde{y}).\, [\widetilde{v}/\widetilde{x}]P : [\widetilde{v}/\widetilde{x}]z.\, (((\langle\widetilde{v}/\widetilde{x}\rangle A_2)\uparrow_{\{\widetilde{y}\}})$$

As in the previous case we push the substitution out in the process and the behavior to get, as required,

$$\Gamma \triangleright [\widetilde{v}/\widetilde{x}](z(\widetilde{y}).\, P) : \langle\widetilde{v}/\widetilde{x}\rangle(z.\, (A_2\uparrow_{\{\widetilde{y}\}})).$$ □

**Lemma B.3** (Subject-reduction)**.**

(1) *If* $\Gamma \triangleright P : A$ *and* $P \preceq Q$ *then* $\Gamma \triangleright Q : A$.

(2) *(Subject-reduction) If* $P \xrightarrow{L} P'$ *and* $\Gamma \triangleright P : A$ *then* $A \overset{L}{\Longrightarrow} A'$ *and* $\Gamma \triangleright P' : A'$ *for some* $A'$.

*Proof.* **Part 1.** By induction on the derivation of $P \preceq Q$. Most cases use Lemma A.2. The case for $(\nu x)\, P|Q \preceq (\nu x)\, (P|Q)$ uses Lemma B.1. The only interesting case is that for $(\mathfrak{N}^{\Phi} x)P|Q \preceq (\mathfrak{N}^{\Phi} x)(P|Q)$ with $x \notin \mathbf{FV}(Q)$. The judgement $\Gamma \triangleright (\mathfrak{N}^{\Phi} x)P|Q : A$ must have been inferred from

$$\begin{array}{c}\Gamma, x : \mathbf{res} \triangleright P : A_3 \qquad \mathbf{traces}_x(A_3) \subseteq \Phi \qquad A_3\uparrow_{\{x\}} \le A_1 \\ \Gamma \triangleright Q : A_2 \qquad A_1 | A_2 \le A\end{array}$$

From these and Lemma B.1, we infer

$$\Gamma, x : \mathbf{res} \triangleright P|Q : A_3|A_2.$$

By alpha-renaming assume $x \notin \mathbf{FV}(A_2)$. By Lemmas A.6.6 and A.6.11 we get $(A_3|A_2){\downarrow_{\{x\}}} \approx A_3{\downarrow_{\{x\}}}|A_2{\downarrow_{\{x\}}} \leq A_3{\downarrow_{\{x\}}}$, and then by Lemma A.7.1 we get $\mathbf{traces}_x(A_3|A_2) \subseteq \mathbf{traces}_x(A_3)$, and so $\mathbf{traces}_x(A_3|A_2) \subseteq \Phi$. This gives

$$\Gamma \triangleright (\mathfrak{N}^\Phi x)(P|Q) : (A_3|A_2){\uparrow_{\{x\}}}.$$

Finally $(A_3|A_2){\uparrow_{\{x\}}} \leq A_3{\uparrow_{\{x\}}}|A_2 \leq A_1 \,|\, A_2 \leq A$. This gives as required

$$\Gamma \triangleright (\mathfrak{N}^\Phi x)(P|Q) : A.$$

**Extended case only:** Just replace **traces** with **etraces** in the above reasoning.

**Part 2.** By induction on the derivation of $P \xrightarrow{L} P'$. We show main cases. The other cases are straightforward.

- Case (R-Com): We are given

$$\Gamma \triangleright \overline{x}\langle\widetilde{v}\rangle.\, P_1 \,|\, x(\widetilde{y}).\, P_2 : A.$$

This must have been deduced from

$$\Gamma \triangleright \overline{x}\langle\widetilde{v}\rangle.\, P_1 : A_1$$
$$\Gamma \triangleright x(\widetilde{y}).\, P_2 : A_2$$
$$A_1|A_2 \leq A. \tag{B.1}$$

$\Gamma \triangleright \overline{x}\langle\widetilde{v}\rangle.\, P_1 : A_1$ and $\Gamma \triangleright x(\widetilde{y}).\, P_2 : A_2$ must have been deduced from

$$\Gamma \triangleright P_1 : A_3 \tag{B.2}$$
$$\Gamma \triangleright x : \mathbf{chan}\langle(\widetilde{y} : \widetilde{\sigma})A_4\rangle \tag{B.3}$$
$$\Gamma \triangleright v_i : \sigma_i \tag{B.4}$$
$$\overline{x}.\,(\langle\widetilde{v}/\widetilde{y}\rangle A_4 \,|\, A_3) \leq A_1 \tag{B.5}$$

and

$$\Gamma, \widetilde{y} : \widetilde{\sigma} \triangleright P_2 : A_5 \tag{B.6}$$
$$\Gamma \triangleright x : \mathbf{chan}\langle(\widetilde{y} : \widetilde{\sigma})A_4\rangle$$
$$A_5{\downarrow_{\{\widetilde{y}\}}} \leq A_4 \tag{B.7}$$
$$x.\,(A_5{\uparrow_{\{\widetilde{y}\}}}) \leq A_2 \tag{B.8}$$

respectively. We must show $A \Rightarrow A'$ and $\Gamma \triangleright P_1|[\widetilde{v}/\widetilde{y}]P_2 : A'$ for some $A'$. We pick some $A'$ such that $A \Rightarrow A'$ and $A' \geq \langle\widetilde{v}/\widetilde{y}\rangle A_4|A_3|A_5{\uparrow_{\{\widetilde{y}\}}}$. The existence of such $A'$ is guaranteed by $A \geq \overline{x}.\,(\langle\widetilde{v}/\widetilde{y}\rangle A_4|A_3)|x.\,(A_5{\uparrow_{\{\widetilde{y}\}}}) \longrightarrow \langle\widetilde{v}/\widetilde{y}\rangle A_4|A_3|A_5{\uparrow_{\{\widetilde{y}\}}}$, which follows from (B.1) and (B.5) and (B.8), and the definition of the subtyping relation (Definition 3.5). It remains to prove $\Gamma \triangleright P_1|[\widetilde{v}/\widetilde{y}]P_2 : A'$. We start with the judgment (B.6),

$$\Gamma, \widetilde{y} : \widetilde{\sigma} \triangleright P_2 : A_5.$$

By Lemma B.2.2,

$$\Gamma \triangleright [\widetilde{v}/\widetilde{y}]P_2 : \langle\widetilde{v}/\widetilde{y}\rangle A_5.$$

Hence

$$\Gamma \triangleright P_1|[\widetilde{v}/\widetilde{y}]P_2 : A_3|\langle\widetilde{v}/\widetilde{y}\rangle A_5.$$

Therefore, the required result $\Gamma \triangleright P_1 \,|\, [\widetilde{v}/\widetilde{y}]P_2 : A'$ follows by (T-Sub), if we show $A_3 \,|\, \langle\widetilde{v}/\widetilde{y}\rangle A_5 \;\leq\; A'$. It follows by:

$$\begin{array}{rcll} A_3 \,|\, \langle\widetilde{v}/\widetilde{y}\rangle A_5 & \leq & A_3 \,|\, \langle\widetilde{v}/\widetilde{y}\rangle(A_5{\downarrow_{\{\widetilde{y}\}}} \,|\, A_5{\uparrow_{\{\widetilde{y}\}}}) & \text{(Lemma A.7.5)} \\ & \leq & A_3 \,|\, \langle\widetilde{v}/\widetilde{y}\rangle(A_5{\downarrow_{\{\widetilde{y}\}}}) \,|\, \langle\widetilde{v}/\widetilde{y}\rangle(A_5{\uparrow_{\{\widetilde{y}\}}}) & \text{(Lemma A.5.4)} \\ & \leq & A_3 \,|\, \langle\widetilde{v}/\widetilde{y}\rangle(A_5{\downarrow_{\{\widetilde{y}\}}}) \,|\, A_5{\uparrow_{\{\widetilde{y}\}}} & \text{(Lemma A.5.10)} \\ & \leq & A_3 \,|\, \langle\widetilde{v}/\widetilde{y}\rangle A_4 | A_5{\uparrow_{\{\widetilde{y}\}}} & \text{(assumption B.7 above)} \\ & \leq & A' & \text{(the definition of } A'\text{).} \end{array}$$

- Case (R-Acc): We are given $\Gamma \triangleright \mathbf{acc}_\xi(x).P_1 : A$. This must have been derived from
  - $\Gamma \triangleright P_1 : A_1$
  - $\Gamma \triangleright x : \mathbf{res}$
  - $x^\xi.A_1 \;\leq\; A$.

  We have to show that
  - $\Gamma \triangleright P_1 : A'$
  - $A \stackrel{x^\xi}{\Longrightarrow} A'$.

  Let $A'$ be a behavioral type that satisfies $A \stackrel{x^\xi}{\Longrightarrow} A'$ and $A' \;\geq\; A_1$. Such $A'$ is guaranteed to exist by $A \;\geq\; x^\xi.A_1 \stackrel{x^\xi}{\longrightarrow} A_1$. Then, $\Gamma \triangleright P_1 : A'$ follows from $\Gamma \triangleright P_1 : A_1$ and $A' \;\geq\; A_1$.
- Case (R-NewR1): We are given $\Gamma \triangleright (\mathfrak{N}^\Phi x)P_1 : A$ This must have been derived from
  - $\Gamma, x : \mathbf{res} \triangleright P_1 : A_1$
  - $\mathbf{traces}_x(A_1) \subseteq \Phi$
  - $A \;\geq\; A_1{\uparrow_{\{x\}}}$.

  We have to show that there exists $A'$ such that
  - $\Gamma \triangleright (\mathfrak{N}^{\Phi^{-\xi}} x)P'_1 : A'$
  - $A \Longrightarrow A'$

  where $P_1 \stackrel{x^\xi}{\longrightarrow} P'_1$.

  By the induction hypothesis, there exists $A'_1$ that satisfies $\Gamma, x : \mathbf{res} \triangleright P'_1 : A'_1$ and $A_1 \stackrel{x^\xi}{\Longrightarrow} A'_1$. Using (Tr-Project), we get $A_1{\downarrow_{\{x\}}} \stackrel{x^\xi}{\Longrightarrow} A'_1{\downarrow_{\{x\}}}$. So, from the definition of **traces** and $\mathbf{traces}_x(A_1) \subseteq \Phi$, we get $\mathbf{traces}_x(A'_1) \subseteq \Phi^{-\xi}$. By using (T-NewR), we get $\Gamma \triangleright (\mathfrak{N}^{\Phi^{-\xi}} x)P'_1 : A'_1{\uparrow_{\{x\}}}$.

  It remains to show there exists $A'$ such that $A'_1{\uparrow_{\{x\}}} \;\leq\; A'$ and $A \Longrightarrow A'$. That follows from $A \;\geq\; A_1{\uparrow_{\{x\}}} \Longrightarrow A'_1{\uparrow_{\{x\}}}$. Here, the latter relation follows from $A_1 \stackrel{x^\xi}{\Longrightarrow} A'_1$ and rule (Tr-Exclude).

  **Extended case only:** Just replace **traces** with **etraces** in the above reasoning.
- Case (R-SP): This follows immediately from Part 1 and the induction hypothesis. □

## Appendix C. Proofs of the Lemma for Theorem 5.8

This section gives a proof of the lemma "**Disabled**" used in the proof of Theorem 5.8.

**Lemma C.1** (Disabled)**.** *If well_annotated*$(P)$ *and* $\Gamma \triangleright_{pl} P : A$ *with* ***bool*** $\notin$ *codom*$(\Gamma)$*, then* $P \not\longrightarrow$ *implies disabled*$(A, S)$ *for any* $S$*.*

*Proof.* We first note that *well_annotated*$(P)$ and $P \not\longrightarrow$ imply $\neg$*active*$(P)$ by the definition of *well_annotated*$(P)$. So, it is sufficient to show (i)$\Gamma \triangleright_{pl} P : A$, (ii)$P \not\longrightarrow$, (iii)$\neg$*active*$(P)$, and (iv) **bool** $\notin$ *codom*$(\Gamma)$ imply *disabled*$(A, S)$ for any $S$. We prove this by induction on the derivation of $\Gamma \triangleright_{pl} P : A$, with case analysis on the last rule.

- Case (T-Zero): In this case, $A = \mathbf{0}$, so we have *disabled*$(A, S)$ for any $S$.
- Case (T-Out): In this case, $P = \overline{x}_t\langle\widetilde{v}\rangle.\, P_1$ and $A = \overline{x}_t.\,(\langle\widetilde{v}/\widetilde{y}\rangle A_1 \mid A_2)$. Since $\neg active(P)$, $t = \emptyset$. So, we have *disabled*$(A, S)$ for any $S$.
- Case (T-In): In this case, $P = x_t(\widetilde{y}).\, P_1$ and $A = x_t.\,(A_2{\uparrow_{\{\widetilde{y}\}}})$. Since $\neg active(P)$, $t = \emptyset$. So, we have *disabled*$(A, S)$ for any $S$.
- Case (T-Par): In this case, $P = P_1 \mid P_2$ and $A = A_1 \mid A_2$ with $\Gamma \triangleright_{pl} P_1 : A_1$ and $\Gamma \triangleright_{pl} P_2 : A_2$. Note that $P \not\longrightarrow$ implies $P_1 \not\longrightarrow$ and $P_2 \not\longrightarrow$. $\neg active(P)$ implies $\neg active(P_1)$ and $\neg active(P_2)$. So, by the induction hypothesis, we get *disabled*$(A_1, S)$ and *disabled*$(A_2, S)$ for any $S$, which implies *disabled*$(A, S)$.
- Case (T-Rep): In this case, $P = *P_1$ and $A = *A_1$, with $\Gamma \triangleright_{pl} P_1 : A_1$. $\neg active(P)$ and $P \not\longrightarrow$ imply $\neg active(P_1)$ and $P_1 \not\longrightarrow$. So, by the induction hypothesis, we get *disabled*$(A_1, S)$ for any $S$, which also implies *disabled*$(A, S)$ as required.
- Case (T-If): This case cannot happen; by the condition (iv), $P$ must be of the form **if true then** $P_1$ **else** $P_2$ or **if false then** $P_1$ **else** $P_2$ , which contradicts with $P \not\longrightarrow$.
- Case (T-New): In this case, $P = (\nu x)\, P_1$, $A = (\nu x)\, A_2$, and $\Gamma, x : \mathbf{chan}\langle(\widetilde{y} : \widetilde{\sigma})A_1\rangle \triangleright P_1 : A_2$. $\neg active(P)$ and $P \not\longrightarrow$ imply $\neg active(P_1)$ and $P_1 \not\longrightarrow$. So, by the induction hypothesis, we get *disabled*$(A_2, S)$ for any $S$. By the definition of *disabled*$(\cdot, S)$, we get *disabled*$(A, S)$.
- Case (T-Acc): This case cannot happen, since $P$ must be of the form $\mathbf{acc}_\xi(x).P_1$, which contradicts with $P \not\longrightarrow$.
- Case (T-NewR): Similar to the case for (T-New).
- Case (T-Sub): $\Gamma \triangleright_{pl} P : A$ must be derived from $\Gamma \triangleright_{pl} P : A'$ for some $A' \leq A$. By the induction hypothesis, for any $S$, we get *disabled*$(A', S)$. By the condition $A' \leq A$, we have *disabled*$(A, S)$ for any $S$. □

## Appendix D. Computing a Basis of Behavioral Type

This section is an appendix for Section 4.3.1. Let $A$ be a behavioral type of the form $(\nu\widetilde{y})\, B$, where $B$ does not contain any $\nu$-prefix. Such $A$ can be obtained by pushing all the $\nu$-prefixes out to the top-level, as described in Section 4.3.1. We show how to compute a basis of $A$ below.

The constructor $\cdot\uparrow_S$ can be eliminated by running the algorithm $\mathit{ElimUp}^{\emptyset,\emptyset}(B,\emptyset)$ below.

$$
\begin{array}{l}
\mathit{ElimUp}^{F,D}(\mathbf{0},S) = \mathbf{0}\\
\mathit{ElimUp}^{F,D}(\alpha,S) =\\
\quad \begin{cases} A & \text{if } F(\alpha,S)=A\\ \mu\beta.\mathit{ElimUp}^{F\{(\alpha,S)\mapsto\beta\},D}(D(\alpha),S) & \text{if } (\alpha,S)\notin \mathit{dom}(F)\end{cases}\\
\mathit{ElimUp}^{F,D}(l.A,S) = (l\backslash S).\mathit{ElimUp}^{F,D}(A,S)\\
\mathit{ElimUp}^{F,D}(A_1 \,|\, A_2,S) = \mathit{ElimUp}^{F,D}(A_1,S) \,|\, \mathit{ElimUp}^{F,D}(A_2,S)\\
\mathit{ElimUp}^{F,D}(A_1 \oplus A_2,S) =\\
\quad \mathit{ElimUp}^{F,D}(A_1,S) \oplus \mathit{ElimUp}^{F,D}(A_2,S)\\
\mathit{ElimUp}^{F,D}(*A,S) = *\mathit{ElimUp}^{F,D}(A,S)\\
\mathit{ElimUp}^{F,D}(\langle\widetilde{y}/\widetilde{x}\rangle A,S) = \mathit{ElimUp}^{F,D}(A,\{z \mid [\widetilde{y}/\widetilde{x}]z\in S\})\\
\mathit{ElimUp}^{F,D}(\mu\alpha.A,S) = \mu\alpha.\mathit{ElimUp}^{F\{(\alpha,S)\mapsto\alpha\},D\{\alpha\mapsto A\}}(A,S)\\
\mathit{ElimUp}^{F,D}(A\uparrow_{S_1},S) = \mathit{ElimUp}^{F,D}(A,S\cup S_1)\\
\mathit{ElimUp}^{F,D}(A\downarrow_{S_1},S) = \mathit{ElimUp}^{F,D}(A,S)\downarrow_{S_1}
\end{array}
$$

Here, $l\backslash S$ is $\tau$ if $\mathit{target}(l)\subseteq S$ and $l$ otherwise. $D$ keeps recursive definitions and $F$ is a cache for avoiding repeated computation. If $A$ does not contain $\nu$-prefixes, $\mathit{ElimUp}^{\emptyset,\emptyset}(B,\emptyset)$ always terminates since $S$ can range over a finite set (which is the powerset of $\mathbf{FV}(B)$). The constructor $\cdot\downarrow_S$ can be removed in the same manner.

We can further eliminate the renaming constructor $\langle\widetilde{y}/\widetilde{x}\rangle$ by using the following algorithm.

$$
\begin{array}{l}
\mathit{ElimRen}^{F,D}(\mathbf{0},\theta) = \mathbf{0}\\
\mathit{ElimRen}^{F,D}(\alpha,\theta) =\\
\quad \begin{cases} A & \text{if } F(\alpha,\theta)=A\\ \mu\beta.\mathit{ElimRen}^{F\{(\alpha,\theta)\mapsto\beta\},D}(D(\alpha),\theta) & \text{if } (\alpha,\theta)\notin \mathit{dom}(F)\end{cases}\\
\mathit{ElimRen}^{F,D}(l.A,\theta) = \theta l.\mathit{ElimRen}^{F,D}(A,\theta)\\
\mathit{ElimRen}^{F,D}(A_1 \,|\, A_2,\theta) = \mathit{ElimRen}^{F,D}(A_1,\theta) \,|\, \mathit{ElimRen}^{F,D}(A_2,\theta)\\
\mathit{ElimRen}^{F,D}(A_1 \oplus A_2,\theta) =\\
\quad \mathit{ElimRen}^{F,D}(A_1,\theta) \oplus \mathit{ElimRen}^{F,D}(A_2,\theta)\\
\mathit{ElimRen}^{F,D}(*A,\theta) = *\mathit{ElimRen}^{F,D}(A,\theta)\\
\mathit{ElimRen}^{F,D}(\langle\widetilde{y}/\widetilde{x}\rangle A,\theta) = \mathit{ElimRen}^{F,D}(A,\theta\circ[\widetilde{y}/\widetilde{x}])\\
\mathit{ElimRen}^{F,D}(\mu\alpha.A,\theta) = \mu\alpha.\mathit{ElimRen}^{F\{(\alpha,\theta)\mapsto\alpha\},D\{\alpha\mapsto A\}}(A,\theta)
\end{array}
$$

By applying the above algorithms to $A = (\nu\widetilde{y})\,B$, we obtain an equivalent type $A' = (\nu\widetilde{y})\,B'$, where $B'$ does not contain any $\nu$-prefixes, $\cdot\downarrow_S$, $\cdot\uparrow_S$, or $\langle\widetilde{y}/\widetilde{x}\rangle$. So, only elements of $\mathbf{Atoms}(B')$ defined below (modulo folding/unfolding of recursive types) can appear in transitions of $B$. So, $(\{\widetilde{y}\},\mathbf{Atoms}(B'))$ forms a basis of $A$.

**Definition D.1.** Let $A$ be a behavioral type that does not contain any $\nu$-prefix, $\cdot\downarrow_S$, $\cdot\uparrow_S$, or $\langle\widetilde{y}/\widetilde{x}\rangle$. The set of *atoms* $\mathbf{Atoms}(A)$ is the least set that satisfies the following conditions.

$$
\begin{array}{l}
\mathbf{Atoms}(l.A) \supseteq \{l.A\}\cup\mathbf{Atoms}(A)\\
\mathbf{Atoms}(A_1 \,|\, A_2) \supseteq \mathbf{Atoms}(A_1)\cup\mathbf{Atoms}(A_2)\\
\mathbf{Atoms}(A_1\oplus A_2) \supseteq \{A_1\oplus A_2\}\cup\mathbf{Atoms}(A_1)\cup\mathbf{Atoms}(A_2)\\
\mathbf{Atoms}(*A) \supseteq \{*A\}\cup\mathbf{Atoms}(A)\\
\mathbf{Atoms}(\mu\alpha.A) \supseteq \{\mu\alpha.A\}\cup\mathbf{Atoms}([\mu\alpha.A/\alpha]A)
\end{array}
$$